\documentclass[preprint,aps,prl,superscriptaddress]{revtex4-1}

\usepackage{graphicx}% Include figure files
\usepackage{dcolumn}% Align table columns on decimal point
\usepackage{bm}% bold math
\usepackage{amsmath}
\usepackage{latexsym}
\usepackage{hyperref} % Comment it when submitting.

\newcommand{\rfig}[1]{Fig.~\ref{#1}}

\newcommand{\req}[1]{Eq.~(\ref{#1})}

\makeindex

\begin{document}

\title{V$_5$S$_8$: a Kondo lattice based on intercalation of van der Waals layered transition metal dichalcogenide}

\author{Jingjing Niu}
\altaffiliation{These authors equally contributed to the work.}
\affiliation{State Key Laboratory for Artificial Microstructure and Mesoscopic Physics, Beijing Key Laboratory of Quantum Devices, Peking University, Beijing 100871, China}
\affiliation{Collaborative Innovation Center of Quantum Matter, Beijing 100871, China}
\author{Zhilin Li}
\altaffiliation{These authors equally contributed to the work.}
\affiliation{State Key Laboratory for Artificial Microstructure and Mesoscopic Physics, Beijing Key Laboratory of Quantum Devices, Peking University, Beijing 100871, China}
\affiliation{Collaborative Innovation Center of Quantum Matter, Beijing 100871, China}
\author{Sixian Yang}
\affiliation{CAS Key Laboratory of Quantum Information, Synergetic Innovation Center of Quantum Information and Quantum Physics, University of Science and Technology of China, Hefei 230026, Anhui, China}
\author{Wenjie Zhang}
\affiliation{State Key Laboratory for Artificial Microstructure and Mesoscopic Physics, Beijing Key Laboratory of Quantum Devices, Peking University, Beijing 100871, China}
\affiliation{Collaborative Innovation Center of Quantum Matter, Beijing 100871, China}
\author{Dayu Yan}
\affiliation{Institute of Physics and Beijing National Laboratory for Condensed Matter Physics, Chinese Academy of Sciences, Beijing 100190, China}
\author{Shulin Chen}
\affiliation{Electron Microscopy Laboratory, School of Physics, Peking University, Beijing 100871, China}
\author{Zhepeng Zhang}
\affiliation{Center for Nanochemistry, Beijing National Laboratory for Molecular Sciences, College of Chemistry and Molecular Engineering, Peking University, Beijing 100871, China}
\author{Yanfeng Zhang}
\affiliation{Center for Nanochemistry, Beijing National Laboratory for Molecular Sciences, College of Chemistry and Molecular Engineering, Peking University, Beijing 100871, China}
\author{Xinguo Ren}
\affiliation{CAS Key Laboratory of Quantum Information, Synergetic Innovation Center of Quantum Information and Quantum Physics, University of Science and Technology of China, Hefei 230026, Anhui, China}
\author{Peng Gao}
\affiliation{Collaborative Innovation Center of Quantum Matter, Beijing 100871, China}
\affiliation{Electron Microscopy Laboratory, School of Physics, Peking University, Beijing 100871, China}
\affiliation{International Center for Quantum Materials, School of Physics, Peking University, Beijing 100871, China}
\author{Youguo Shi}
\affiliation{Institute of Physics and Beijing National Laboratory for Condensed Matter Physics, Chinese Academy of Sciences, Beijing 100190, China}
\author{Dapeng Yu}
\affiliation{State Key Laboratory for Artificial Microstructure and Mesoscopic Physics, Beijing Key Laboratory of Quantum Devices, Peking University, Beijing 100871, China}
\affiliation{Collaborative Innovation Center of Quantum Matter, Beijing 100871, China}
\affiliation{Department of Physics, Southern University of Science and Technology of China, Shenzhen 518055, China}
\author{Xiaosong Wu}
\email{xswu@pku.edu.cn}
\affiliation{State Key Laboratory for Artificial Microstructure and Mesoscopic Physics, Beijing Key Laboratory of Quantum Devices, Peking University, Beijing 100871, China}
\affiliation{Collaborative Innovation Center of Quantum Matter, Beijing 100871, China}
\affiliation{Department of Physics, Southern University of Science and Technology of China, Shenzhen 518055, China}

\begin{abstract}
Since the discovery of graphene, a tremendous amount of two dimensional (2D) materials have surfaced. Their electronic properties can usually be well understood without considering correlations between electrons. On the other hand, strong electronic correlations are known to give rise to a variety of exotic properties and new quantum phases, for instance, high temperature superconductivity, heavy fermions and quantum spin liquids. The study of these phenomena has been one of the main focuses of condensed matter physics. There is a strong incentive to introduce electronic correlations into 2D materials. Via intercalating a van der Waals layered compound VS$_2$, we show an emergence of a Kondo lattice, an extensively studied strongly correlated system, by magnetic, specific heat, electrical and thermoelectric transport studies. In particular, an exceptionally large Sommerfeld coefficient, 440 mJ$\cdot$K$^{-2}\cdot$mol$^{-1}$, indicates a strong electron correlation. The obtained Kadowaki-Woods ratio, $2.7\times 10^{-6}$ $\mu\Omega\cdot$cm$\cdot$mol$^2\cdot$K$^2\cdot$mJ$^{-2}$, also supports the strong electron-electron interaction. The temperature dependence of the resistivity and thermopower corroborate the Kondo lattice picture. The intercalated compound is one of a few rare examples of $d$-electron Kondo lattices. We further show that the Kondo physics persists in ultra-thin films. This work thus demonstrates a route to generate strong correlations in 2D materials.
\end{abstract}


\maketitle

The past decade has seen an explosion in research of two dimensional (2D) materials\cite{Geim2013,Bhimanapati2015}. During discovery of a flood of 2D materials, many salient effects in solid, e.g., integer and fractional quantum Hall effect, quantum spin Hall effect, superconductivity and magnetism, have been observed in 2D materials\cite{Wu2018,Lu2015,Huang2017,Gong2017}. Most of the electronic properties of 2D materials can be described by a single-electron picture. On the other hand, strong electron correlations have been one of the most important problems in condensed matter physics and are believed to play a pivotal role in systems, such as high temperature superconductors, heavy fermions, and quantum spin liquids, etc\cite{Lee2006,Stewart1984,Lee2008}. There has been little interaction between these two fields. Very recently, emergence of strong correlations in twisted bilayer graphene has stirred much excitement in both communities\cite{Cao2018,Cao2018a}. The excellent tunability via electrical gating and surface engineering of 2D materials may provide to the experimental investigation of strong correlation phenomena a powerful knob that is not available in research of three dimensional bulk.

2D materials are derived from their three dimensional counterpart, van der Waals layered compounds. Many kinds of atoms and molecules, varying significantly in size and property, can be inserted into the interlayer gap due to weak interlayer coupling. In fact, there have been extensive studies on intercalated graphite and transition metal dichalcogenides (TMD's)\cite{Marseglia1983,Friend1987,Dresselhaus2002}. New electrical, optical and magnetic properties have been introduced to the host material this way. Such intercalation can be achieved in bilayers\cite{Sugawara2011,Kanetani2012}, hence making a new 2D material. Therefore, intercalation is expected to greatly expand the spectrum of properties of 2D materials. A naive idea is to insert local magnetic moments so that they couple to the conduction carriers in the layer, forming a Kondo lattice. This could be a versatile scheme for designing strongly correlated 2D electronic systems.

Here, we demonstrate that the scheme is viable by showing that V$_5$S$_8$, a self-intercalated compound of VS$_2$, is a Kondo lattice. VS$_2$ is a non-magnetic layered TMD\cite{Murphy1977}. V$_5$S$_8$, however, becomes an antiferromagnetic (AFM) metal because of the local moment of the intercalated V ion. In this work, we study the magnetic susceptibility, specific heat, electrical and thermoelectric transport of V$_5$S$_8$. The Sommerfeld coefficient is found to be 440 and 67 mJ$\cdot$K$^{-2}$ per mole of local moments above and below the N\'{e}el temperature, respectively. The exceptionally large coefficients indicate renormalization of the effective electron mass due to strong electron correlation. Both the electrical resistivity and thermopower show characteristics of a Kondo lattice, suggesting that the strong correlation results from the Kondo coupling between itinerant electrons in the plane and intercalated local moments. The low temperature resistivity follows a $T^2$ dependence, consistent with the Fermi liquid effect. The Kadowaki-Woods (KW) ratio is $\sim$0.27$a_0$, indicating strong electron-electron scattering. Our results reveal that V$_5$S$_8$ is a $d$-electron Kondo lattice compound and thus show that intercalation of van der Waals layered materials can be an effective and versatile method to bring new effects, such as strong correlation, into many known 2D materials.

%\section{Experimental techniques}
V$_5$S$_8$ bulk single crystals were grown by a chemical vapor transport method, using vanadium and sulfur powders as precursors and iodine as a transport agent. These species were loaded into a silica ampule under argon. The ampule was then evacuated, sealed and heated gradually in a two-zone tube furnace to a temperature gradient of $1000\ ^\circ$C to $850\ ^\circ$C. After two weeks, single crystals with regular shapes and shiny facets can be obtained. V$_5$S$_8$ thin flakes were grown by a chemical vapour deposition method\cite{Niu2017a}. X-ray experiments on grown crystals indicate a pure V$_5$S$_8$ phase. The crystallographic structure of the single crystal was further confirmed by high-angle annular dark field scanning transmission electron microscope (HAADF-STEM). Transport properties were measured using a standard lock-in method in an OXFORD variable temperature cryostat from 1.5 to 300 K. Heat capacity was measured in a Quantum Design Physical Properties Measurement System. A Quantum Design SQUID magnetometer was employed to measure the magnetic susceptibility. Thermopower of bulk single crystals was measured using a standard four-probe steady-state method with a Chromel/AuFe(0.07\%) thermocouple, while a micro-heater method was used for thin films\cite{Jia2016}.

\begin{figure}[htbp]
\includegraphics[width=0.9\textwidth]{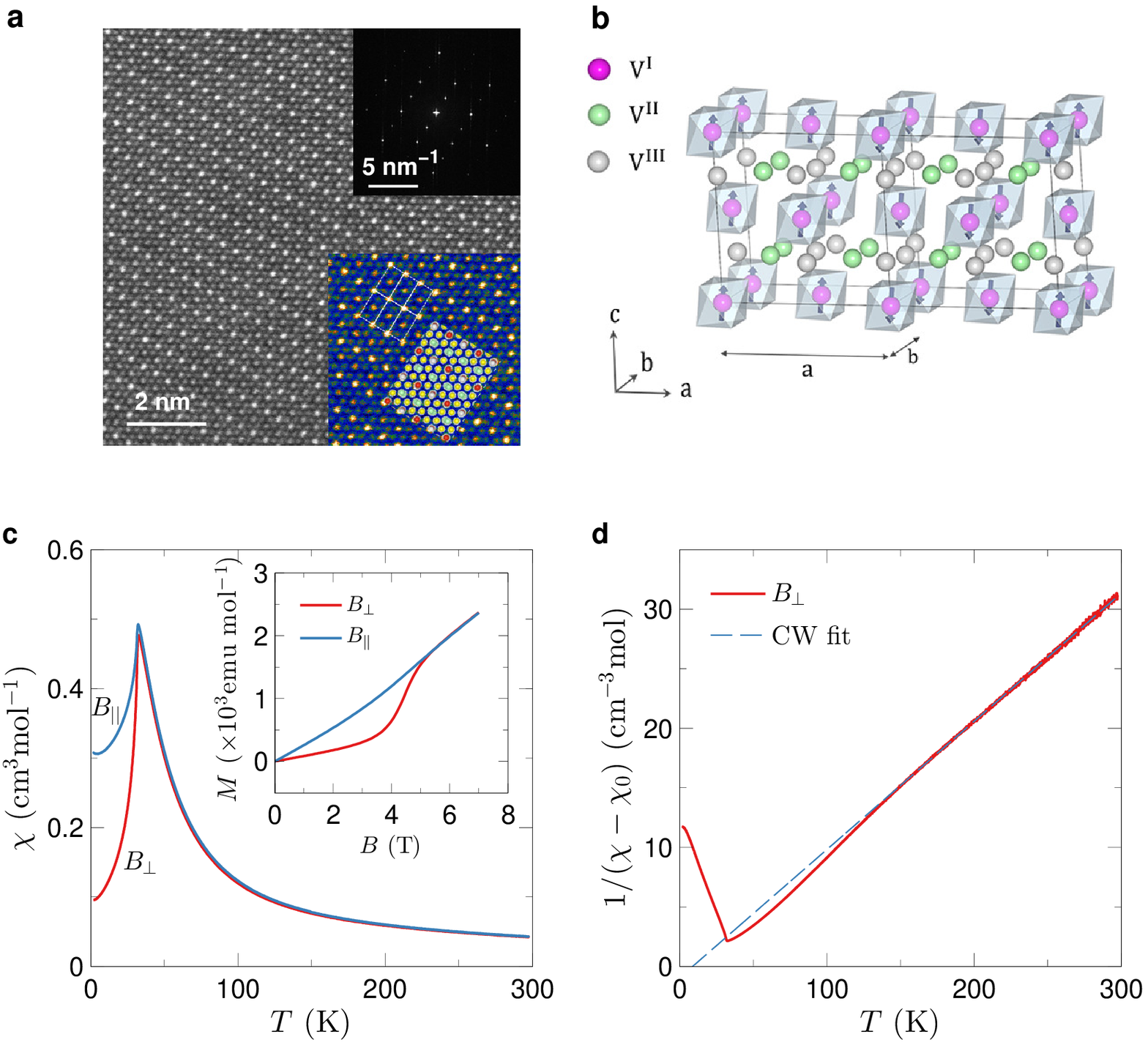}
\centering
\caption{Structural and magnetic properties of V$_5$S$_8$ bulk single crystal. \textbf{a}, HAADF-STEM image. Lower inset, a zoom-in image with the in-plane atomic model. Upper inset, a reduced FFT image. \textbf{b}, the magnetic unit cell with the V$^\mathrm{I}$S$_6$ octahedron. The blue arrows indicate the direction of the magnetic moments on V$^\mathrm{I}$ sites. \textbf{c}, the molar magnetic susceptibility $\chi$ for $B_\perp$ ( $B\perp ab$ plane) and $B_\parallel$ ($B\parallel ab$ plane). The inset shows the low temperature ($T=2$ K) isothermal magnetization curves. \textbf{d}, the inverse magnetic susceptibility $1/(\chi-\chi_0)$ for $B_\perp$. The blue dashed line is a linear fit of the Curie-Weiss law, which gives $\theta=8.8$ K, $\chi_0=0.01$ cm$^3$mol$^{-1}$ and $\mu_{eff}=2.43\mu_\mathrm{B}$ per V$^\mathrm{I}$.}
\label{fig2}
\end{figure}

%\subsection{Structure}
V$_5$S$_8$ can be seen as a van der Waals layered material VS$_2$ self-intercalated with V, V$_{1/4}$VS$_2$. It crystallizes in a monoclinic structure, space group $C2/m$. V atoms lie on three inequivalent sites. Intercalated V atoms take the V$^\mathrm{I}$ site, while V atoms in the VS$_2$ layer take V$^\mathrm{II}$ and V$^\mathrm{III}$ sites. Each V$^\mathrm{I}$ atom is surrounded by six S atoms, forming a distorted octahedron, shown in \rfig{fig2}(b). The resultant crystal field is believed to be intricately related to the local magnetic moment of V atoms\cite{Silbernagel1975,Fujimori1991}. \rfig{fig2}(a) shows a HAADF-STEM image of a single crystal, in which both V and S atoms can be clearly seen, as well as the rectangular arrangement of the intercalated V$^\mathrm{I}$ atoms. Images have been taken at various spots. All of them show high crystallinity and the same lattice orientation, indicating uniform V intercatlation. A zoom-in and color-enhanced image is shown in the lower right of \rfig{fig2}(a), which is in excellent accordance with the in-plane atomic model of V$_5$S$_8$. The lattice constants, $a=11.65$ {\AA} and $b=6.76$ {\AA}, are determined from a Fast Fourier Transformation (FFT) image shown in the upper right of \rfig{fig2}(a).

%\subsection{Magnetic properties}
The magnetic susceptibility of V$_5$S$_8$ displays a paramagnetic behavior which follows the Curie-Weiss (CW) law at high temperatures and undergoes an antiferromagnetic transition at about 32 K, shown in \rfig{fig2}\cite{Nozaki1978}. The high temperature CW behavior suggests presence of local moments. Fitting of the paramagnetic susceptibility to the CW law, $\chi=\chi_0+\frac{C}{T-\theta_\mathrm{CW}}$, yields a positive CW temperature $\theta_\mathrm{CW}\approx 8.8$ K despite the AFM order and an effective magnetic moment of $\mu_\mathrm{eff}=2.43\mu_\mathrm{B}$ per V$_5$S$_8$ formula unit (f.u.). This value is consistent with reported ones, ranging from 2.12 to 2.49 $\mu_\mathrm{B}$\cite{Vries1973,Nishihara1977,Nozaki1978}. It is generally believed that only V$^\mathrm{I}$ ions carry a local magnetic moment\cite{Silbernagel1975,Kitaoka1980,Fujimori1991}. So, the measured moment is close to the theoretical value $2.64\mu_\mathrm{B}$ for V$^{3+}(3d^2)$\cite{Silbernagel1975,Nozaki1978,Katsuta1979}. Below the N\'{e}el temperature, the moments point in the direction of 10.4$^\circ$ away from the $c$ axis toward the $a$ axis. They align as $\uparrow\uparrow\downarrow\downarrow$ along the $a$ axis, while they align ferromagnetically along the $b$ and $c$ axes\cite{Silbernagel1975}. Consequently, the in-plane ($B\parallel ab$ plane) susceptibility is larger than the out-of-plane one. The field dependence of the out-of-plane magnetization displays a sudden change of the slope at about 4.5 T, seen in the inset of \rfig{fig2}(c). This is caused by a metamagnetic spin-flop transition\cite{Nozaki1978}. A neutron scattering study found a moment of 0.7 and 1.5$\mu_\mathrm{B}$ in two single crystals, respectively\cite{Funahashi1981}, while a nuclear magnetic resonance study found a much smaller moment, 0.22 $\mu_\mathrm{B}$\cite{Kitaoka1980}. Taking the measured paramagnetic moment of $2.43\mu_\mathrm{B}$, $S\approx0.8$ is obtained. The moment below $T_\mathrm{N}$ mostly seems smaller than the expected value $2S=1.7$\cite{Nakanishi2000}. This discrepancy has been an unresolved puzzle. We would also like to point out that the susceptibility deviates from the CW law below 140 K. In the following, we are going to show that this puzzle and the deviation can be explained in terms of hybridization of the local $d$-electrons on V$^\mathrm{I}$ with the conduction electrons in the VS$_2$ plane, a correlation effect known as the Kondo effect.

\begin{figure}[htbp]
\includegraphics[width=0.9\textwidth]{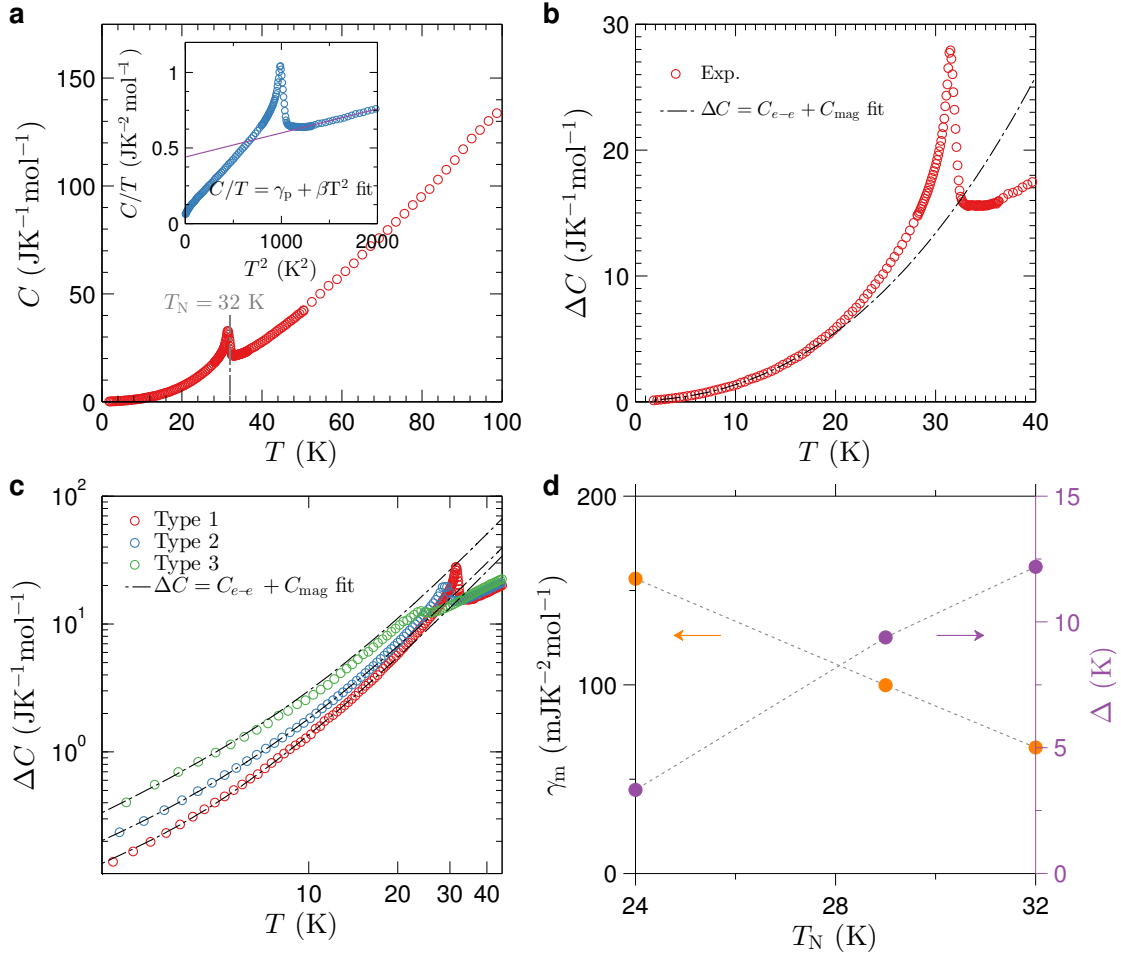}
\centering
\caption{Specific heat. \textbf{a}, the specific heat $C$ as a function of temperature. The inset shows a $C/T$ versus $T^2$ plot. The solid line is a linear fit, which yields $\gamma_\mathrm{p}=440$ mJ K$^{-2}$ mol$^{-1}$. \textbf{b}, the low-$T$ specific heat $\Delta C$ after subtracting the lattice contribution $\beta T^3$ according to the high-$T$ fit. The dashed line represents a fit to the \req{CT}. \textbf{c}, the non-lattice part of specific heat $\Delta C$ for three types of samples with different $T_\mathrm{N}$. The dashed lines are the fits using \req{CT}. \textbf{d}, the fitting parameters as a function of $T_\mathrm{N}$.}
\label{fig3}
\end{figure}

%\subsection{Specific heat}
There have already been indications for interactions between the localized $d$-electrons and the conduction electrons in this system. Anomalous Hall effect, due to skew scattering of conduction electrons off from local moments, has been observed in V$_5$S$_8$\cite{Niu2017a}. Photoemission-spectroscopy study has shown both local-moment-like and bandlike features for V $3d$ electrons. To understand the discrepancy, it was thus postulated that the $3d$ electron that provides the local moment becomes partially itinerant at low temperatures\cite{Fujimori1991}. Recently, in a similar compound, VSe$_2$ with dilute V intercalation, the Kondo effect has been observed\cite{Barua2017}. In V$_5$S$_8$, where local moments of V$^\mathrm{I}$ atoms arrange in a periodic array, a Kondo lattice is naturally anticipated. As a result, the effective mass of the conduction electron will be substantially enhanced by the electron correlation. We have carried out specific heat measurements for V$_5$S$_8$. The data of a 3.18 mg bulk single crystal are presented in \rfig{fig3}(a). A sizeable jump at $T_\mathrm{N}=32$ K manifests the AFM transition. In the inset of \rfig{fig3}(a), $C(T)/T$ is plotted as a function of $T^2$. A linear dependence is observed above $T_\mathrm{N}$. By a linear fit to $C/T=\gamma_\mathrm{p}+\beta T^2$, a very large electronic Sommerfeld coefficient, $\gamma_\mathrm{p}=440$ mJ$\cdot$K$^{-2}$ per mole of V$^\mathrm{I}$ atoms, is obtained. The value is comparable to those in $f$-electron heavy-fermion systems that exhibit a magnetic order at low temperatures, e.g., 148 mJ$\cdot$K$^{-2}\cdot$mol$^{-1}$ in Ce$_2$Rh$_3$Ge$_5$\cite{Hossain1999} and 504 mJ$\cdot$K$^{-2}\cdot$mol$^{-1}$ in U$_2$Zn$_{17}$\cite{Fisk1986}. Coincidentally, $\gamma=420$ mJ$\cdot$K$^{-2}\cdot$mol$^{-1}$ in a vanadium oxide, LiV$_2$O$_4$, which is the first $d$-electron heavy fermion metal\cite{Kondo1997}. The observed large Sommerfeld coefficient provides direct evidence for the mass enhancement in V$_5$S$_8$.

Below $T_\mathrm{N}$, taking into account the spin wave contribution, the non-lattice part of the specific heat can be expressed as
\begin{equation}\label{CT}
\Delta C=\gamma_\mathrm{m}T+A_C\Delta^4\sqrt{\frac{T}{\Delta}}\mathrm{e}^{-\Delta/T}\left[1+\frac{39}{20}\left(\frac{T}{\Delta}\right)+\frac{51}{32}\left(\frac{T}{\Delta}\right)^2\right],
\end{equation}
where $\gamma_\mathrm{m}$ represents the electron contribution in the magnetically ordered state, $\Delta$ is the spin-wave gap\cite{Pikul2003,Zhou2018,Continentino2001,Szlawska2012}. In \rfig{fig3}(b), we plot the low-$T$ $\Delta C$ curve obtained by subtracting the lattice contribution ($\beta T^3$) according to the high-$T$ fit, \emph{i.e.}, $\Delta C=C-\beta T^3$. The data below 20 K can be well fitted to \req{CT}, producing $\gamma_\mathrm{m}=67$ mJ$\cdot$K$^{-2}$mol$^{-1}$ and $\Delta=12.2$ K. Though much smaller than $\gamma_\mathrm{p}$, $\gamma_\mathrm{m}$ is still comparable to some strongly correlated materials\cite{Hossain1999,Li2004}. The suppression of the Sommerfeld coefficient by magnetic ordering is typical in heavy fermion systems that order magnetically at low temperatures\cite{Hossain1999,Fisk1986}. This is because of the competition between the Kondo coupling, which results in mass enhancement, and the Ruderman-Kittel-Kasuya-Yosida (RKKY) interaction, which promotes a magnetic order\cite{Doniach1977}.

The competition is also reflected in the observed negative correlation between $\gamma_\mathrm{m}$ and $T_\mathrm{N}$. \rfig{fig3}(c) compares the non-lattice part of the low-$T$ specific heat $\Delta C$ for three types of V$_5$S$_8$ samples. These samples display variation of $T_\mathrm{N}$, possibly due to different intercalation level\cite{Nozaki1978}. As $T_\mathrm{N}$ decreases, the spin-wave gap $\Delta$ gradually is reduced, but $\gamma_\mathrm{m}$ increases from 67 to 156 mJ$\cdot$K$^{-2}$mol$^{-1}$. Such a negative correlation is illustrated in the Doniach phase diagram of heavy fermion systems, indicating approaching to the quantum critical point\cite{Doniach1977}.

Given the key effect of intercalated V atoms in introducing the Kondo effect, it would be informative to compare the intercalate V$_{1/4}$VS$_2$ and the host compound VS$_2$. However, it is challenging to grow VS$_2$ because of self intercalation and contradicting properties have been observed so far\cite{Gauzzi2014}. Consequently, we have measured the specific heat of VSe$_2$ instead, an isoelectronic and isostructural compound of VS$_2$, shown in the Supporting Materials. The Sommerfeld coefficient is found to be 46 mJ$\cdot$K$^{-2}$ per mol of V$_4$Se$_8$, smaller than that of V$_5$S$_8$. This difference offers additional evidence for enhanced electronic correlation by the Kondo coupling to local moments.

To get an idea of the strength of the electron correlation, we estimate the effective quasiparticle mass enhanced by the strong correlation by comparing the experimental Sommerfeld coefficient to that calculated from the Kohn-Sham model to density functional theory (DFT)\cite{Kohn1965}. Within such a non-interacting electron model, the Sommerfeld coefficient can be estimated as $\gamma=\frac{1}{3}\pi^2 k_\mathrm{B}^2N(\epsilon_\mathrm{F})$, where $k_\mathrm{B}$ is the Boltzmann constant and $N(\epsilon_\mathrm{F})$ is the density of states (DOS) per f.u. at the Fermi level $\epsilon_\textrm{F}$. Our DFT calculation for the antiferromagetic phase of V$_5$S$_8$ yields an electronic DOS of 6.6 states/eV/f.u., which translates to a Sommerfeld coefficient of only 16.8 mJ$\cdot$K$^{-2}\cdot$mol$^{-1}$. To understand the difference between the experimental value and the calculated one here, it is necessary to first take into account the electron-phonon coupling which can enhance the Sommerfeld coefficient by a factor of (1+$\lambda_\mathrm{ep})$, where $\lambda_{\mathrm{ep}}$ is the mass enhancement factor due to the electron-phonon coupling\cite{BECK1970}. A reasonable estimate of $\lambda_\mathrm{ep}$ is 1.19\cite{Savrasov1996}, which was obtained in V metal. This leads to an enhanced Sommerfeld coefficient of 36.8 mJ$\cdot$K$^{-2}\cdot$mol$^{-1}$. Thus, one can see that, even after accounting for the electron-phonon coupling effect, the theoretical value estimated from a quasiparticle picture is still a factor of 2 smaller than the experimental one. We attribute the remaining discrepancy to the mass enhancement effect due to strong electron correlations. In the antiferromagnetic phase, the electron correlation gives rise to an enhanced effective mass of 1.82 $m_\mathrm{e}$ with $m_\mathrm{e}$ being the bare electron mass. In the paramagnetic phase, this mass enhancement effect is much more pronounced, where the effective mass is estimated to be 11.4 $m_e$. The details of our DFT calculation for the electronic DOS are included in the supplementary materials.

%\subsection{Temperature-dependent resistivity behavior}

\begin{figure}[htbp]
\includegraphics[width=1\textwidth]{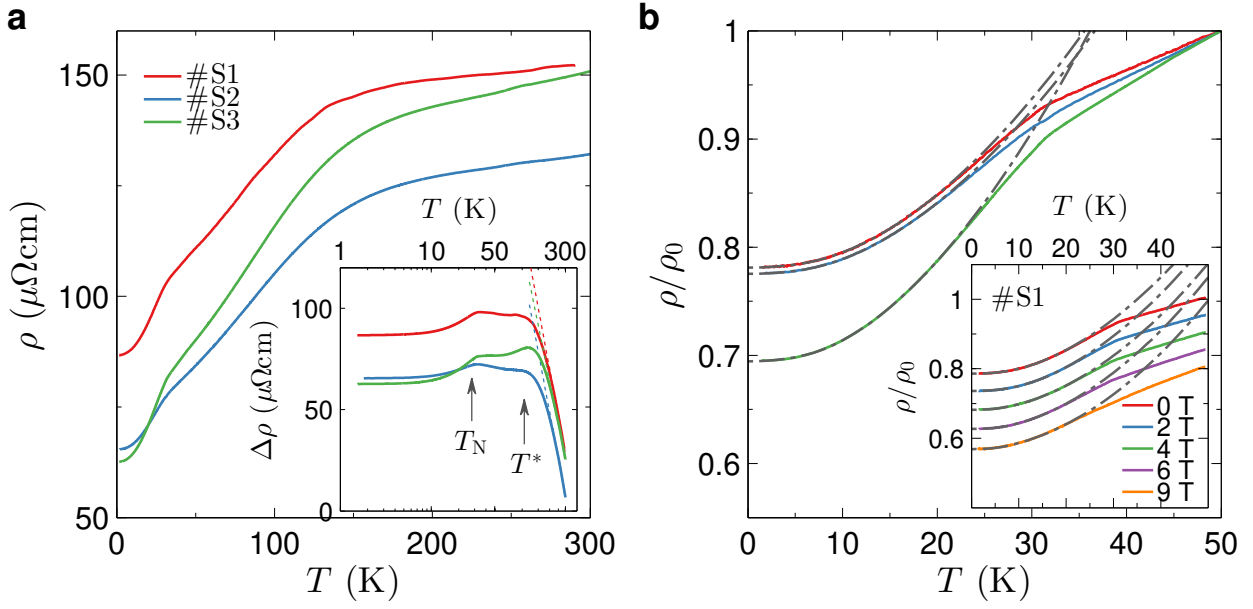}
\centering
\caption{Temperature dependence of resistivity. \textbf{a}, the temperature-dependent resistivity of three samples. The inset illustrates the high-$T$ resistivity after subtracting the non-magnetic $\rho$ of VSe$_2$ from that of V$_5$S$_8$. The dotted lines are the $-\ln T$ fits. \textbf{b}, the normalized low-temperature resistivity $\rho/\rho(T=50\ \mathrm{K})$. The dashed lines indicate the fits using \req{RT}. The inset shows the low-$T$ resistivity $\rho/\rho(T=50\ \mathrm{K})$ (vertically shifted for clarity.) for Sample $\#$S1 at different magnetic fields ($B \perp ab$ plane).}
\label{fig1}
\end{figure}

The temperature dependent resistivity of many metallic Kondo lattice materials exhibits characteristic features, such as a maximum, stemming from the Kondo scattering\cite{Fisk1986,Pikul2003}. \rfig{fig1} shows the resistivity for three V$_5$S$_8$ single crystals, \#S1, \#S2, and \#S3. At 32 K, the resistivity exhibits a kink, which results from the antiferromagnetic transition. Above this temperature, there is an apparent hump at about 140 K, in stark contrast to a linear dependence commonly seen in metals. In fact, VSe$_2$ displays a typical metallic resistivity linear in $T$. We tentatively subtract the resistivity of VSe$_2$ from V$_5$S$_8$ to highlight the effect of intercalated atoms. As shown in the inset of \rfig{fig1}a, the broad maximum is evident. Similar features have been observed in Kondo lattices and believed to originate from a combined effect of the Kondo coupling and the crystal field\cite{Fisk1986,Hossain1999,Hossain2000,Pikul2003}. On the high temperature side of the maximum, the resistivity roughly follows a $-\ln T$ dependence, consistent with the effect of incoherent Kondo scattering. The resistivity maximum is at about $T^*=140$ K, suggesting that the crystal field splitting is qualitatively of the order of 140 K.

We now turn to the low temperature resistivity in the AFM state. In the magnetically ordered state, the strong decrease of the resistivity below $T_\mathrm{N}$ is caused by the reduction of spin-disorder scattering. In this case, the resistivity consists of both the electronic contribution and the magnon scattering term, and takes the form of
\begin{equation}\small\label{RT}
 \rho(T)=\rho_0+A T^2+C\Delta^2\sqrt{\frac{T}{\Delta}}\mathrm{e}^{-\Delta/T}\left[1+\frac{2}{3}\left(\frac{T}{\Delta}\right)+\frac{2}{15}\left(\frac{T}{\Delta}\right)^2\right],
\end{equation}
where $\rho_0$ is the residual resistivity, the $AT^2$ term represents the Fermi liquid contribution, the last term is associated with the spin wave, and $\Delta$ is the spin wave gap\cite{Pikul2003,Zhou2018,Continentino2001,Szlawska2012}. The resistivity below 20 K can be well described by \req{RT} (see \rfig{fig1}b). The fitting parameters $A$ and $\Delta$ are $\sim$0.012 $\mu\Omega\cdot$cm$\cdot$K$^{-2}$ and $\sim28$ K, respectively. Moreover, $A$ is found to be nearly independent of the magnetic field, which is consistent with the Fermi liquid contribution\cite{Raquet2002,Madduri2017}. On the other hand, with increasing $B$, $\Delta$ decreases gradually from 30 to 15 K\cite{Mentink1996,Jobiliong2005}.

In strongly correlated systems, it has been found that the Kadowaki¨CWoods ratio, $r_\mathrm{KW}$=$A/\gamma^2$, is significantly enhanced, around $a_0=1.0\times10^{-5}$ $\mu\Omega\cdot$cm(mol$\cdot$K/mJ)$^2$\cite{Kadowaki1986}. Using the low-$T$ Sommerfeld coefficient $\gamma=67$ mJ$\cdot$K$^{-2}$mol$^{-1}$ and $A=0.012$  $\mu\Omega\cdot$cm, we obtain $r_\mathrm{KW}=0.27a_0$. This value is similar to those in strongly correlated systems such as LiV$_2$O$_4$, V$_2$O$_3$, Sr$_2$RuO$_4$ and Na$_{0.7}$CoO$_2$, etc\cite{McWhan1973,Maeno1997,Urano2000,Miyake1989,Li2004}.

%\subsection{Thermoelectric power}

\begin{figure}[htbp]
\includegraphics[width=0.6\textwidth]{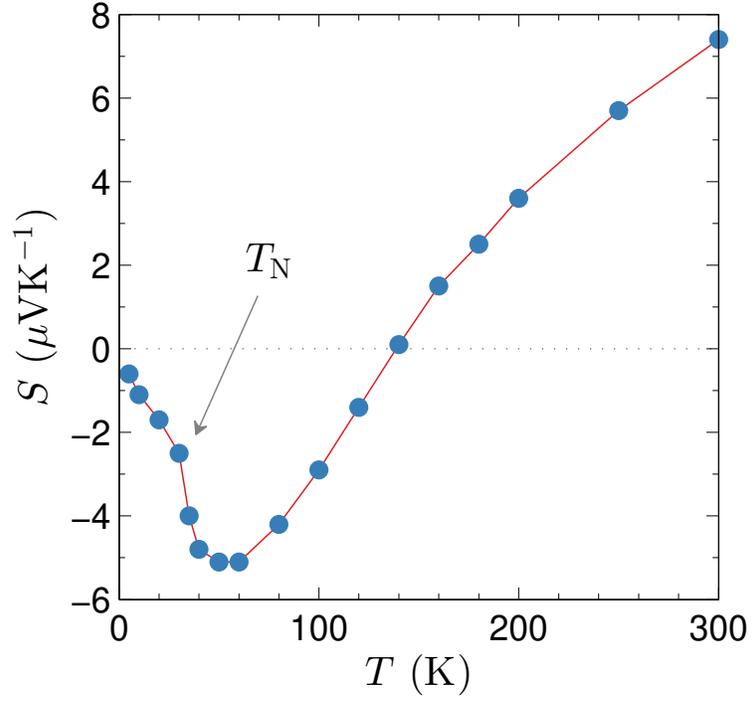}
\centering
\caption{Temperature-dependent thermopower $S$. $S$ changes its sign at about 140 K and displays a negative minimum at $60$ K.}
\label{thermopower}
\end{figure}

The thermoelectric properties of heavy fermion compounds share some common features\cite{Zlatic2003}. \rfig{thermopower} shows the temperature-dependent thermopower $S$ for V$_5$S$_8$. Instead of a linear $T$ dependence as expected for ordinary metals, $S$ shows a sign change at about 140 K, as well as a negative $S$ minimum around 60 K. A change of sign is generally associated with a change of carrier type. However, this explanation is inconsistent with hole conduction inferred from Hall in the whole temperature range (see the Supplementary Material). In a Kondo lattice system, the interplay between the Kondo and crystal field effects gives rise to a broad peak in thermopower $S$ at high temperatures. More prominently, with decreasing temperature, $S$ changes its sign at $T=\alpha T_\mathrm{K}$, where roughly $\alpha=2.5$-10. After that, $S$ displays an extremum and may change sign again in some compounds\cite{Jaccard1992,Zlatic2003,Hodovanets2015,Ren2016}. Our data agree with some of these essential features, e.g., a sign change and a negative peak. It is worth mentioning that the temperature of 140 K, at which $S$ changes its sign, is very close to $T^*$ obtained from the resistivity maximum and deviation of the magnetic susceptibility from the CW law.

%\subsection{Discussion}

Based on these observations, we conclude that V$_5$S$_8$ is a $d$-electron Kondo lattice compound. Under this picture, the magnetic susceptibility can now be understood. The deviation from the CW law beginning at 140 K results from the Kondo coupling and the crystal field effect, which has been seen in other heavy fermion compounds\cite{Hossain2000,Hossain1999,Pikul2003,Fan2004,Szlawska2012}. The reduction of the magnetic local moment at low temperatures is due to the Kondo screening, which, though strongly suppressed, persists in the AFM state\cite{Fisk1986,Si2010}.

%\subsection{Transport properties for thin films}

\begin{figure}[htbp]
\includegraphics[width=1\textwidth]{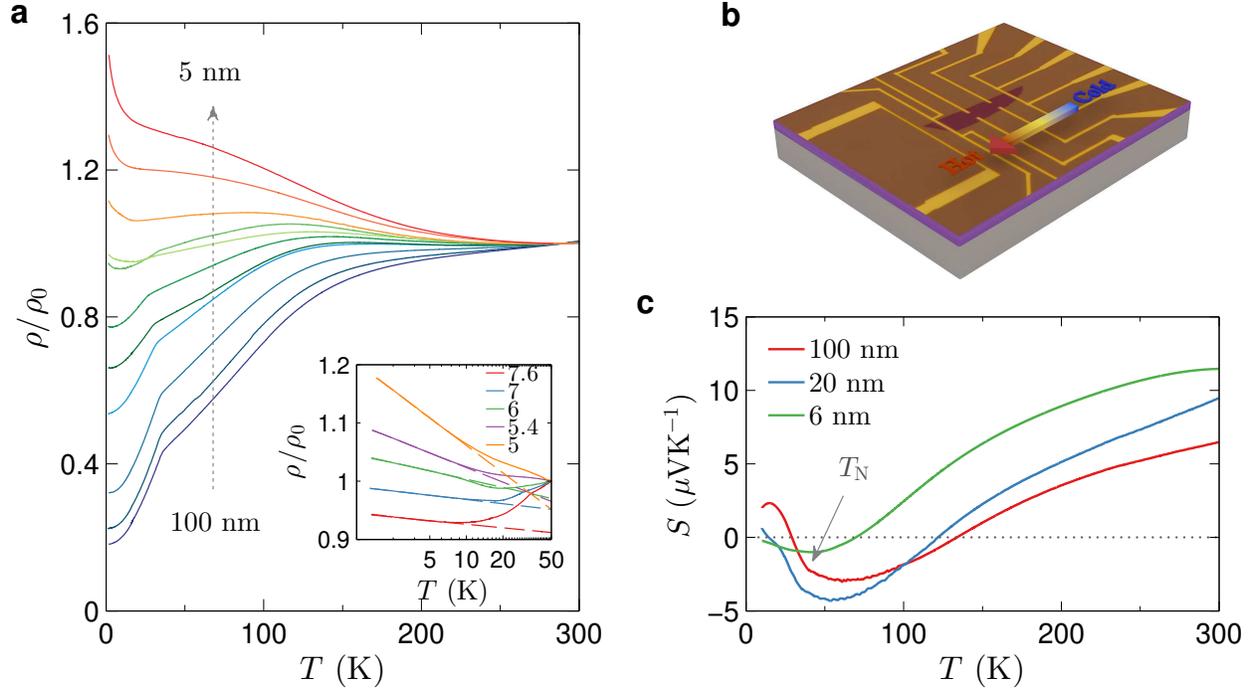}
\centering
\caption{Electrical and thermoelectric transport of V$_5$S$_8$ thin films. \textbf{a}, the temperature dependent resistivity for samples with different thicknesses. The inset shows the low temperature data and $-\ln T$ fits for thin samples. \textbf{b}, a typical device for resistivity $\rho$ and thermopower $S$ measurements.  \textbf{c}, the temperature dependent thermopower for thin films.}
\label{fig4}
\end{figure}

Finally, we investigate the evolution of the correlation effect with reducing thickness. Since it is difficult to measure the specific heat of thin layers, we focus on the resistivity and thermopower and monitor how those features that are relevant to the electron correlation change, as shown in \rfig{fig4}. As samples become thinner, the broad maximum of resistivity shifts towards low temperatures, implying reduction of the crystal field splitting, see \rfig{fig4}(a). The main features in thermopower, e.g., the sign change and the negative peak, remain, shown in \rfig{fig4}(c), though the sign change temperature is suppressed. The same trend takes place in $T_\mathrm{N}$. Below a critical thickness of 5 nm, AFM disappears and a very weak ferromagnetic order emerges\cite{Niu2017a}. One intriguing feature is that the positive temperature coefficient of resistivity below $T_\mathrm{N}$ suddenly turns negative when the thickness is reduced to 7.6 nm (see the inset of \rfig{fig4}(a)). With further reducing thickness, a $-\ln T$ dependence in resistivity develops at low temperatures. It is not clear whether the disappearance of the metallic resistivity reflects a beakdown of Fermi liquid behavior or a metal-to-insulator transition. Another possibility would be an appearance of the second incoherent Kondo scattering contribution due to the crystal field effect\cite{Hossain1999,Hossain2000,Huo2002}. Further investigation are needed. Nevertheless, all these data indicate that the Kondo physics still plays a key role in ultra-thin samples.

%\section{Conclusion}
Our experiments strongly suggest that the intercalated material V$_5$S$_8$ is a $d$-electron Kondo lattice compound. The Kondo physics persists in nano-thick films. The results have not only discovered a 2D strongly correlated material, but open a door to bring tremendous possibilities into 2D material research by intercalation. A recent work has demonstrated that the intercalation can be extended to heterointerfaces, which will further expand the scope of the method\cite{Bediako2018}.

%\section*{Method}

\begin{acknowledgements}
This work was supported by National Key Basic Research Program of China (No. 2013CBA01603, No. 2016YFA0300600, and No. 2016YFA0300903) and NSFC (Project No. 11574005, No. 11774009, No. 11222436 and No. 11574283).
\end{acknowledgements}

% When submitting, the following line needs to be replaced by the content of the .bbl file produced by bibtex.
%\bibliography{VS2}

\clearpage

%%%%%%%%%% Merge with supplemental materials %%%%%%%%%%
\pagebreak
\widetext
\begin{center}
\textbf{\large Supplemental Materials: V$_5$S$_8$: a Kondo lattice based on intercalation of van der Waals layered transition metal dichalcogenide}
\end{center}
%%%%%%%%%% Merge with supplemental materials %%%%%%%%%%
%%%%%%%%%% Prefix a "S" to all equations, figures, tables and reset the counter %%%%%%%%%%
\setcounter{equation}{0}
\setcounter{figure}{0}
\setcounter{table}{0}
\setcounter{page}{1}
\makeatletter
\renewcommand{\theequation}{S\arabic{equation}}
\renewcommand{\thefigure}{S\arabic{figure}}
\renewcommand{\bibnumfmt}[1]{[S#1]}
\renewcommand{\citenumfont}[1]{S#1}
%%%%%%%%%% Prefix a "S" to all equations, figures, tables and reset the counter %%%%%%%%%%

\begin{figure}[htbp]
\includegraphics[width=0.6\textwidth]{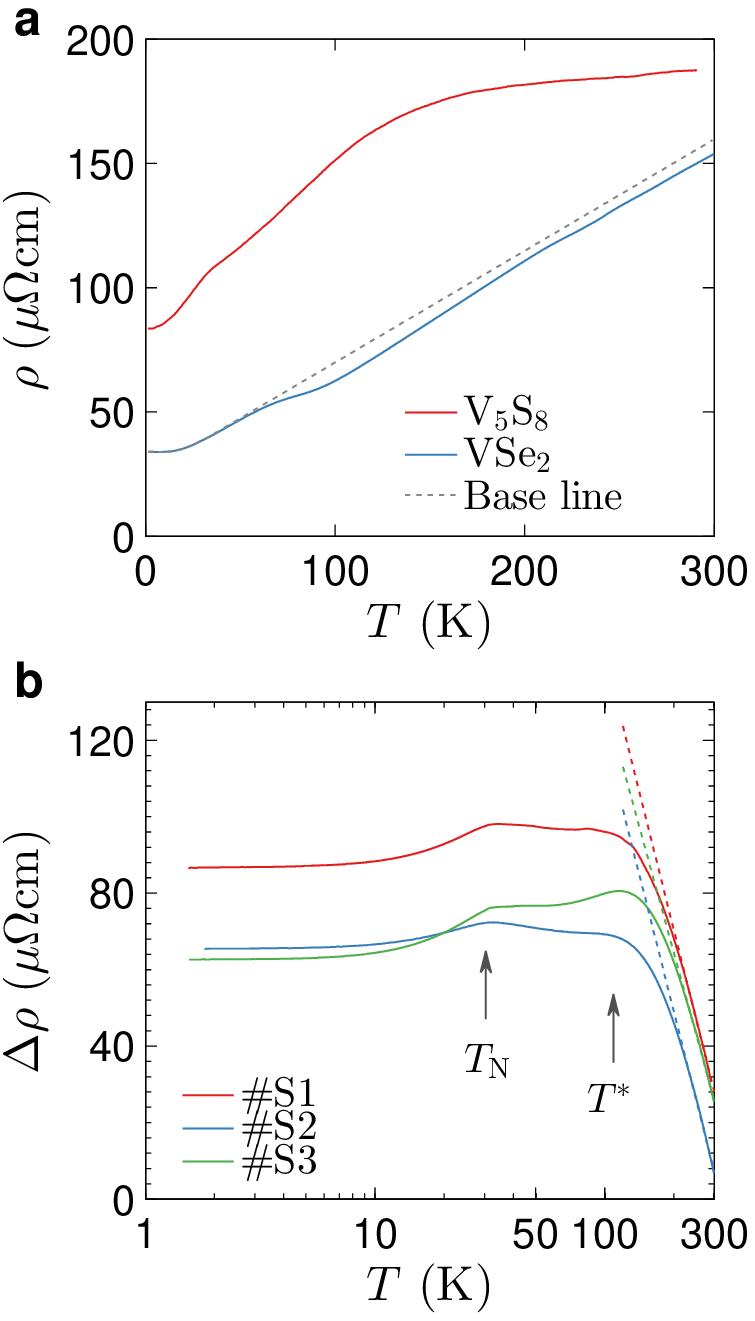}
\centering
\caption{\textbf{a} Comparison of the temperature dependent resistivity $\rho(T)$ between V$_5$S$_8$ and VSe$_2$. $\rho$ versus $T$. VSe$_2$ displays a linear-$T$ dependence, except for an anomaly at $\sim$ 90 K, which is due to an charge density wave transition. The dotted line is the baseline subtracted from the resistivity of V$_5$S$_8$ so as to highlight the contribution of the intercalation, $\Delta \rho$. \textbf{b} $\Delta \rho$ as a function of temperature.}
\label{figs1}
\end{figure}

\newpage

\begin{figure}[htbp]
\includegraphics[width=0.95\textwidth]{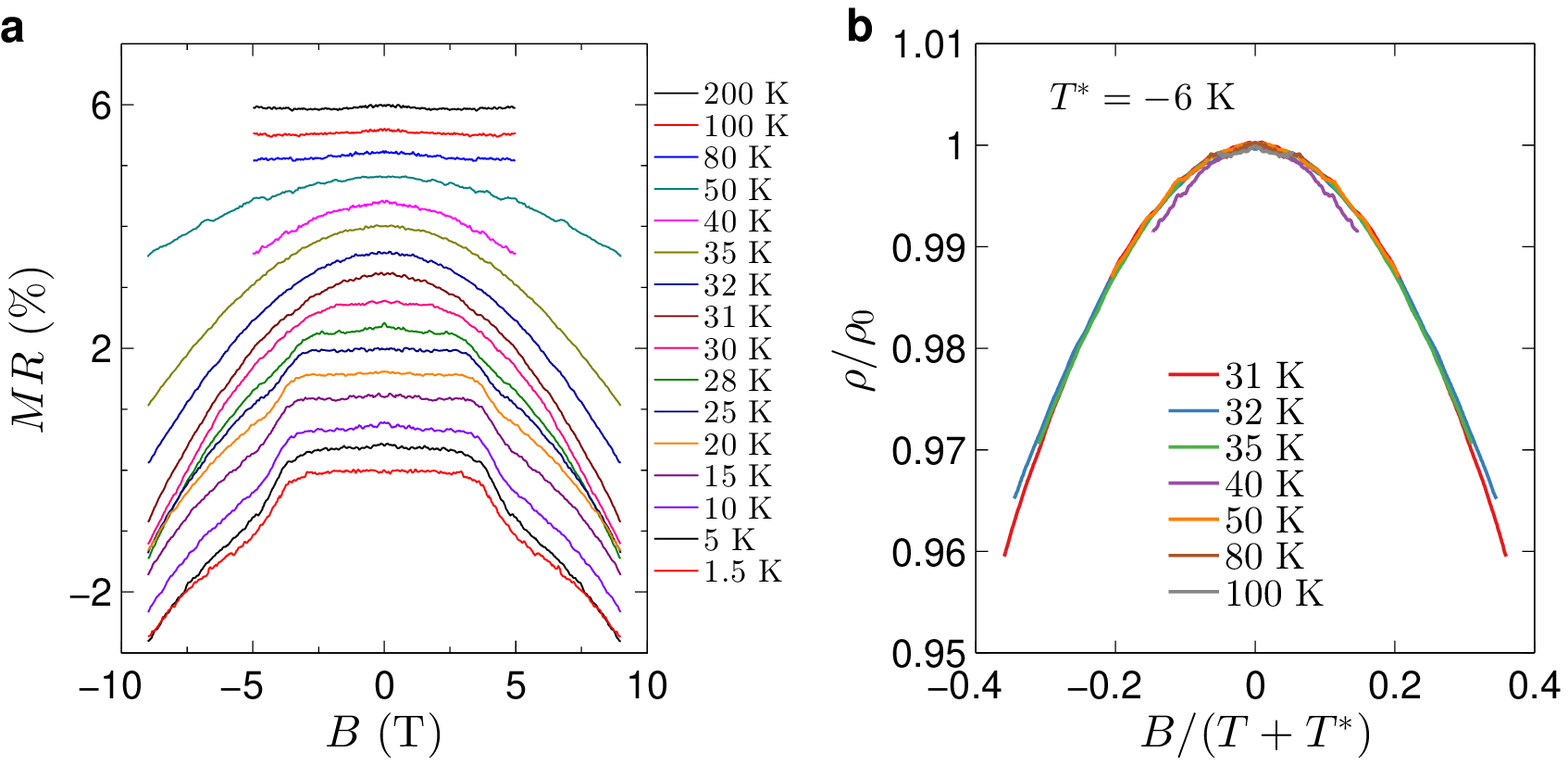}
\centering
\caption{Scaling of magnetoresistance for sample $\#S1$. \textbf{a}, Magnetoresistance $MR$ at various temperatures. The curves are vertically shifted for clarity. \textbf{b}, Normalized resistivity versus $B/(T+T^*)$, where $T^*$ is a scaling parameter.}
\label{figs2}
\end{figure}

We have performed detailed magnetoresistance measurements of V$_5$S$_8$ bulk single crystal. \rfig{figs2}(a) shows the magnetoresistance $MR=\frac{\rho(B)-\rho(0)}{\rho(0)}\times 100\%$ at different temperatures. Above $T_\mathrm{N}$, $MR$ follows a $B^2$ dependence, consistent with the spin fluctuation scattering\cite{Niu2017}. Moreover, $MR$ data collapse onto a single curve when the field is scaled by temperature, $T+T^*$, as shown in \rfig{figs2}(b). It is known that in a Kondo impurity model, the magnetoresistance follows the Schlottmann's relation $\frac{\rho(B)}{\rho(0)}=f[B/(T+T^*)]$\cite{Schlottmann1989a}. The best scaling yields a characteristic temperature of $T^*=-6$ K. The negative sign suggests ferromagnetic correlations in the antiferromagnetic ground state\cite{Hossain2000a,Pikul2003a,Zhou2018b}. This corroborates with the positive Currie-Weiss temperature.

\newpage

\begin{figure}[htbp]
\includegraphics[width=0.95\textwidth]{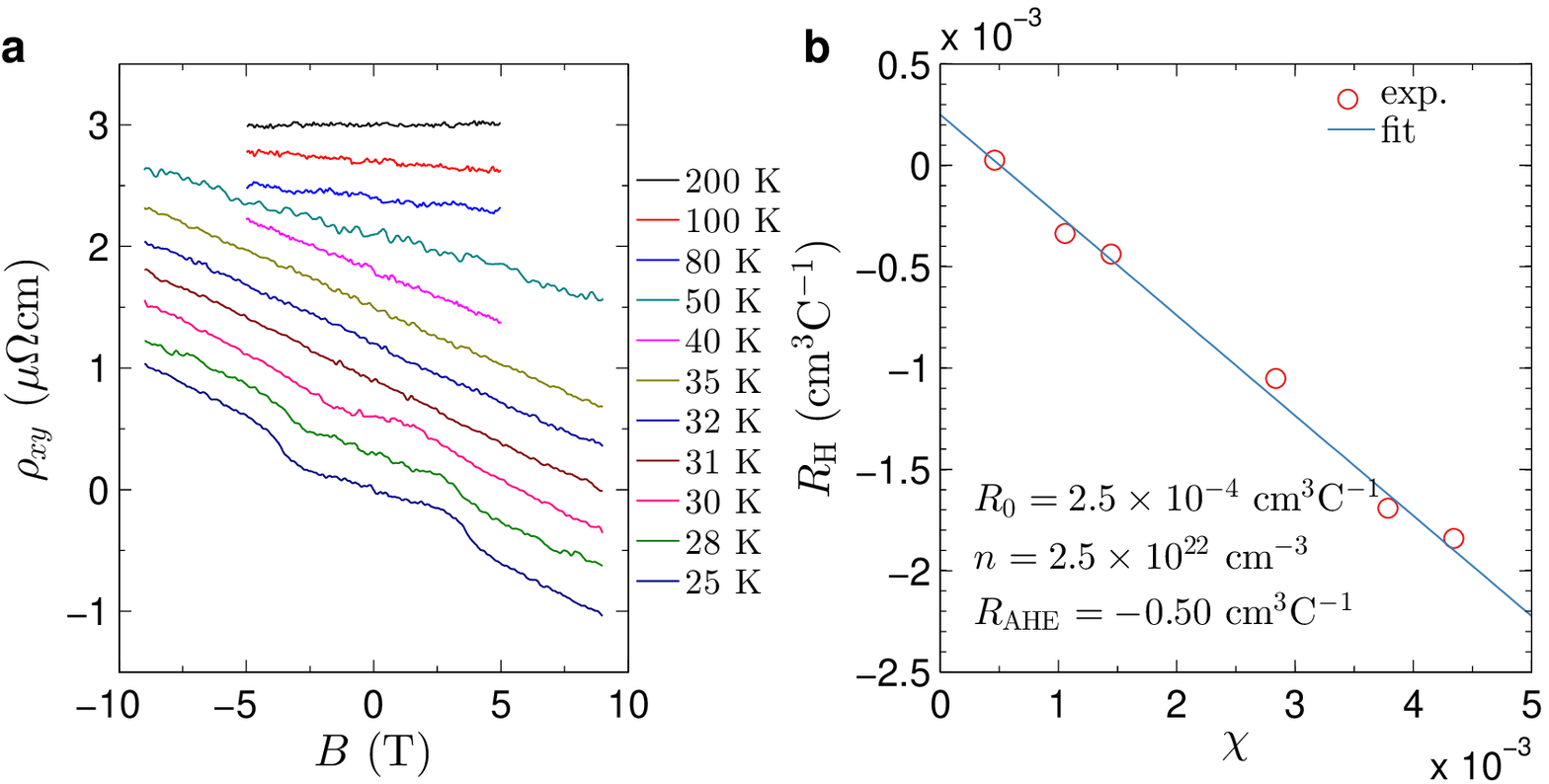}
\centering
\caption{Anomalous Hall effect and the carrier density of V$_5$S$_8$. \textbf{a}, Hall resistivity $\rho_{xy}$ (vertically shifted for clarity) at different temperatures. \textbf{b}, Linear relation between the Hall coefficient $R_\text{H}$ and the magnetic susceptibility $\chi$ above $T_\mathrm{N}$. The blue line is a linear fit.}
\label{figs3}
\end{figure}

We studied the evolution of the Hall resistivity $\rho_{xy}$ with temperature, as shown in \rfig{figs2}(a). Here, $\rho_{xy}$ includes two contributions, \emph{i.e.}, $\rho_ {xy}(B)=R_0 B+R_\text{AHE}\mu_0 M$, where $R_0$ and $R_\text{AHE}$ are the ordinary and anomalous Hall coefficients, respectively, and $M$ is the magnetization, $\mu_0$ the vacuum permeability. So, the Hall coefficient can be expressed as $R_\text{H}=\rho_{xy}/B=R_0 +R_\text{AHE}\mu_0 M/B$,  namely, $R_\text{H}\propto \chi$. We plot $R_\text{H}$ versus $\chi$ at different temperatures above $T_\text{N}$ and find a good linear relation (See \rfig{figs3}(b)). The intercept of the linear fit gives a positive value of $R_0=2.5\times10^{-4} $ cm$^3$ C$^{-1}$, indicating a temperature independent holes carrier concentration of $n=2.5\times10^{22}$ cm$^{-3}$.

\newpage

\begin{figure}[htbp]
\includegraphics[width=0.9\textwidth]{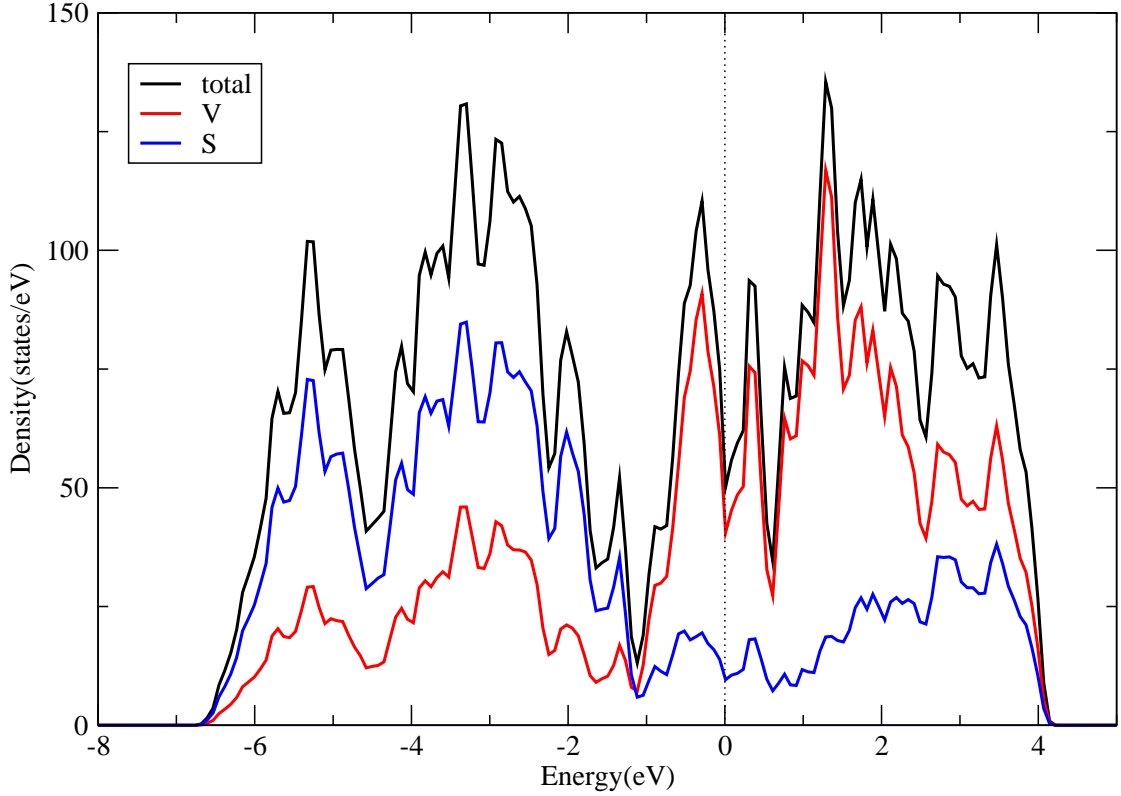}
\centering
\caption{Calculated total (black lines) and species-projected density of states (V: red lines; S: blue lines) of V$_5$S$_8$ bulk. The dash line at the 0 eV represents the Fermi level.}
\label{fig:totalDOS}
\end{figure}

\begin{figure}[htbp]
\includegraphics[width=0.9\textwidth]{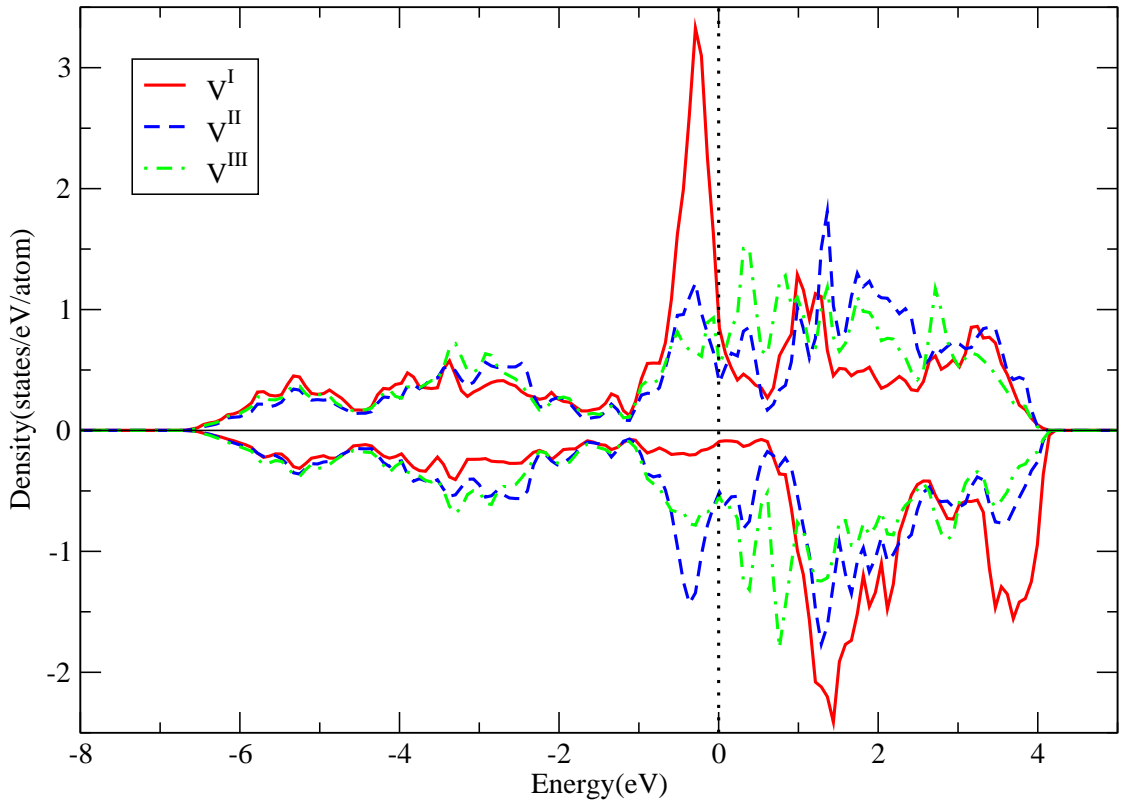}
\centering
\caption{The atom-projected density of states(PDOS) of different Vanadium atoms for V$^\mathrm{I}$ (solid red curve), V$^\mathrm{II}$ (dashed blue curve) and V$^\mathrm{III}$ (dot-dashed green curve) in V$_5$S$_8$ bulk.}
\label{fig:projectedDOS}
\end{figure}

We carried out density functional theory (DFT) calculations using the Perdew-Burke-Ernzerhof (PBE)\cite{Perdew1996a} generalized gradient approximation as implemented in the all-electron first-principles code package \textit{Fritz Haber Institute ab initio molecular simulations} (FHI-aims) package\cite{Blum2009a}. In accord with the experimental findings, the monoclinic crystal structure with antiferromagnetic ordering of the V$^\mathrm{I}$ atoms was used in our calculations. We employed an extended unit cell containing 8 formula unit cells (104 atoms in total), and a $6\times 3 \times 3$ $\bf{k}$ grid (with the $\Gamma$ point included) for the Brillouin zone sampling. The FHI-aims ``light" setting for the numerical grid integration and numerical atomic basis sets ($5s4p2d1f$ for V and $4s3p1d$ for S) were used in the calculations. The unit cell geometry and atomic positions were fully relaxed, with resultant lattice parameters of $a=22.62$ \AA, $b=6.62$ \AA, $c=11.37$ \AA, and $\alpha=\gamma= 90^\circ$, $\beta=91.7^\circ$. The calculated electronic density of states (DOS) are presented in \rfig{fig:totalDOS}, where one can see that the DOS at the Fermi level is dominated by contributions from V. The DOS value of 53.0 states/eV for the calculated supercell corresponds to $N(\epsilon_\mathrm{F})=6.6$ states/eV per formula unit cell. In \rfig{fig:projectedDOS} the projected DOSs of individual V species are presented, where the V$^\mathrm{I}$ atoms yield a pronounced peak just below the Fermi level, and contribute a major part of the spin polarization. A Mulliken charge analysis indicates that the local mangetic moments from V$^\mathrm{I}$, V$^\mathrm{II}$, and V$^\mathrm{III}$ atoms are respectively 1.93, 0.15 and 0.04 $\mu_\mathrm{B}$.

%\bibliography{VS2}

\begin{thebibliography}{60}%
\makeatletter
\providecommand \@ifxundefined [1]{%
 \@ifx{#1\undefined}
}%
\providecommand \@ifnum [1]{%
 \ifnum #1\expandafter \@firstoftwo
 \else \expandafter \@secondoftwo
 \fi
}%
\providecommand \@ifx [1]{%
 \ifx #1\expandafter \@firstoftwo
 \else \expandafter \@secondoftwo
 \fi
}%
\providecommand \natexlab [1]{#1}%
\providecommand \enquote  [1]{``#1''}%
\providecommand \bibnamefont  [1]{#1}%
\providecommand \bibfnamefont [1]{#1}%
\providecommand \citenamefont [1]{#1}%
\providecommand \href@noop [0]{\@secondoftwo}%
\providecommand \href [0]{\begingroup \@sanitize@url \@href}%
\providecommand \@href[1]{\@@startlink{#1}\@@href}%
\providecommand \@@href[1]{\endgroup#1\@@endlink}%
\providecommand \@sanitize@url [0]{\catcode `\\12\catcode `\$12\catcode
  `\&12\catcode `\#12\catcode `\^12\catcode `\_12\catcode `\%12\relax}%
\providecommand \@@startlink[1]{}%
\providecommand \@@endlink[0]{}%
\providecommand \url  [0]{\begingroup\@sanitize@url \@url }%
\providecommand \@url [1]{\endgroup\@href {#1}{\urlprefix }}%
\providecommand \urlprefix  [0]{URL }%
\providecommand \Eprint [0]{\href }%
\providecommand \doibase [0]{http://dx.doi.org/}%
\providecommand \selectlanguage [0]{\@gobble}%
\providecommand \bibinfo  [0]{\@secondoftwo}%
\providecommand \bibfield  [0]{\@secondoftwo}%
\providecommand \translation [1]{[#1]}%
\providecommand \BibitemOpen [0]{}%
\providecommand \bibitemStop [0]{}%
\providecommand \bibitemNoStop [0]{.\EOS\space}%
\providecommand \EOS [0]{\spacefactor3000\relax}%
\providecommand \BibitemShut  [1]{\csname bibitem#1\endcsname}%
\let\auto@bib@innerbib\@empty
%</preamble>
\bibitem [{\citenamefont {Geim}\ and\ \citenamefont
  {Grigorieva}(2013)}]{Geim2013}%
  \BibitemOpen
  \bibfield  {author} {\bibinfo {author} {\bibfnamefont {A.~K.}\ \bibnamefont
  {Geim}}\ and\ \bibinfo {author} {\bibfnamefont {I.~V.}\ \bibnamefont
  {Grigorieva}},\ }\href {http://dx.doi.org/10.1038/nature12385} {\bibfield
  {journal} {\bibinfo  {journal} {Nature}\ }\textbf {\bibinfo {volume} {499}},\
  \bibinfo {pages} {419} (\bibinfo {year} {2013})}\BibitemShut {NoStop}%
\bibitem [{\citenamefont {Bhimanapati}\ \emph {et~al.}(2015)\citenamefont
  {Bhimanapati}, \citenamefont {Lin}, \citenamefont {Meunier}, \citenamefont
  {Jung}, \citenamefont {Cha}, \citenamefont {Das}, \citenamefont {Xiao},
  \citenamefont {Son}, \citenamefont {Strano}, \citenamefont {Cooper},
  \citenamefont {Liang}, \citenamefont {Louie}, \citenamefont {Ringe},
  \citenamefont {Zhou}, \citenamefont {Kim}, \citenamefont {Naik},
  \citenamefont {Sumpter}, \citenamefont {Terrones}, \citenamefont {Xia},
  \citenamefont {Wang}, \citenamefont {Zhu}, \citenamefont {Akinwande},
  \citenamefont {Alem}, \citenamefont {Schuller}, \citenamefont {Schaak},
  \citenamefont {Terrones},\ and\ \citenamefont {Robinson}}]{Bhimanapati2015}%
  \BibitemOpen
  \bibfield  {author} {\bibinfo {author} {\bibfnamefont {G.~R.}\ \bibnamefont
  {Bhimanapati}}, \bibinfo {author} {\bibfnamefont {Z.}~\bibnamefont {Lin}},
  \bibinfo {author} {\bibfnamefont {V.}~\bibnamefont {Meunier}}, \bibinfo
  {author} {\bibfnamefont {Y.}~\bibnamefont {Jung}}, \bibinfo {author}
  {\bibfnamefont {J.}~\bibnamefont {Cha}}, \bibinfo {author} {\bibfnamefont
  {S.}~\bibnamefont {Das}}, \bibinfo {author} {\bibfnamefont {D.}~\bibnamefont
  {Xiao}}, \bibinfo {author} {\bibfnamefont {Y.}~\bibnamefont {Son}}, \bibinfo
  {author} {\bibfnamefont {M.~S.}\ \bibnamefont {Strano}}, \bibinfo {author}
  {\bibfnamefont {V.~R.}\ \bibnamefont {Cooper}}, \bibinfo {author}
  {\bibfnamefont {L.}~\bibnamefont {Liang}}, \bibinfo {author} {\bibfnamefont
  {S.~G.}\ \bibnamefont {Louie}}, \bibinfo {author} {\bibfnamefont
  {E.}~\bibnamefont {Ringe}}, \bibinfo {author} {\bibfnamefont
  {W.}~\bibnamefont {Zhou}}, \bibinfo {author} {\bibfnamefont {S.~S.}\
  \bibnamefont {Kim}}, \bibinfo {author} {\bibfnamefont {R.~R.}\ \bibnamefont
  {Naik}}, \bibinfo {author} {\bibfnamefont {B.~G.}\ \bibnamefont {Sumpter}},
  \bibinfo {author} {\bibfnamefont {H.}~\bibnamefont {Terrones}}, \bibinfo
  {author} {\bibfnamefont {F.}~\bibnamefont {Xia}}, \bibinfo {author}
  {\bibfnamefont {Y.}~\bibnamefont {Wang}}, \bibinfo {author} {\bibfnamefont
  {J.}~\bibnamefont {Zhu}}, \bibinfo {author} {\bibfnamefont {D.}~\bibnamefont
  {Akinwande}}, \bibinfo {author} {\bibfnamefont {N.}~\bibnamefont {Alem}},
  \bibinfo {author} {\bibfnamefont {J.~A.}\ \bibnamefont {Schuller}}, \bibinfo
  {author} {\bibfnamefont {R.~E.}\ \bibnamefont {Schaak}}, \bibinfo {author}
  {\bibfnamefont {M.}~\bibnamefont {Terrones}}, \ and\ \bibinfo {author}
  {\bibfnamefont {J.~A.}\ \bibnamefont {Robinson}},\ }\href {\doibase
  10.1021/acsnano.5b05556} {\bibfield  {journal} {\bibinfo  {journal} {ACS
  Nano}\ }\textbf {\bibinfo {volume} {9}},\ \bibinfo {pages} {11509} (\bibinfo
  {year} {2015})}\BibitemShut {NoStop}%
\bibitem [{\citenamefont {Wu}\ \emph {et~al.}(2018)\citenamefont {Wu},
  \citenamefont {Fatemi}, \citenamefont {Gibson}, \citenamefont {Watanabe},
  \citenamefont {Taniguchi}, \citenamefont {Cava},\ and\ \citenamefont
  {Jarillo-Herrero}}]{Wu2018}%
  \BibitemOpen
  \bibfield  {author} {\bibinfo {author} {\bibfnamefont {S.}~\bibnamefont
  {Wu}}, \bibinfo {author} {\bibfnamefont {V.}~\bibnamefont {Fatemi}}, \bibinfo
  {author} {\bibfnamefont {Q.~D.}\ \bibnamefont {Gibson}}, \bibinfo {author}
  {\bibfnamefont {K.}~\bibnamefont {Watanabe}}, \bibinfo {author}
  {\bibfnamefont {T.}~\bibnamefont {Taniguchi}}, \bibinfo {author}
  {\bibfnamefont {R.~J.}\ \bibnamefont {Cava}}, \ and\ \bibinfo {author}
  {\bibfnamefont {P.}~\bibnamefont {Jarillo-Herrero}},\ }\href@noop {}
  {\bibfield  {journal} {\bibinfo  {journal} {Science}\ }\textbf {\bibinfo
  {volume} {359}},\ \bibinfo {pages} {76} (\bibinfo {year} {2018})}\BibitemShut
  {NoStop}%
\bibitem [{\citenamefont {Lu}\ \emph {et~al.}(2015)\citenamefont {Lu},
  \citenamefont {Zheliuk}, \citenamefont {Leermakers}, \citenamefont {Yuan},
  \citenamefont {Zeitler}, \citenamefont {Law},\ and\ \citenamefont
  {Ye}}]{Lu2015}%
  \BibitemOpen
  \bibfield  {author} {\bibinfo {author} {\bibfnamefont {J.~M.}\ \bibnamefont
  {Lu}}, \bibinfo {author} {\bibfnamefont {O.}~\bibnamefont {Zheliuk}},
  \bibinfo {author} {\bibfnamefont {I.}~\bibnamefont {Leermakers}}, \bibinfo
  {author} {\bibfnamefont {N.~F.~Q.}\ \bibnamefont {Yuan}}, \bibinfo {author}
  {\bibfnamefont {U.}~\bibnamefont {Zeitler}}, \bibinfo {author} {\bibfnamefont
  {K.~T.}\ \bibnamefont {Law}}, \ and\ \bibinfo {author} {\bibfnamefont
  {J.~T.}\ \bibnamefont {Ye}},\ }\href
  {http://www.sciencemag.org/content/early/2015/11/11/science.aab2277.abstract}
  {\bibfield  {journal} {\bibinfo  {journal} {Science}\ }\textbf {\bibinfo
  {volume} {350}},\ \bibinfo {pages} {1353} (\bibinfo {year}
  {2015})}\BibitemShut {NoStop}%
\bibitem [{\citenamefont {Huang}\ \emph {et~al.}(2017)\citenamefont {Huang},
  \citenamefont {Clark}, \citenamefont {Navarro-Moratalla}, \citenamefont
  {Klein}, \citenamefont {Cheng}, \citenamefont {Seyler}, \citenamefont
  {Zhong}, \citenamefont {Schmidgall}, \citenamefont {McGuire}, \citenamefont
  {Cobden}, \citenamefont {Yao}, \citenamefont {Xiao}, \citenamefont
  {Jarillo-Herrero},\ and\ \citenamefont {Xu}}]{Huang2017}%
  \BibitemOpen
  \bibfield  {author} {\bibinfo {author} {\bibfnamefont {B.}~\bibnamefont
  {Huang}}, \bibinfo {author} {\bibfnamefont {G.}~\bibnamefont {Clark}},
  \bibinfo {author} {\bibfnamefont {E.}~\bibnamefont {Navarro-Moratalla}},
  \bibinfo {author} {\bibfnamefont {D.~R.}\ \bibnamefont {Klein}}, \bibinfo
  {author} {\bibfnamefont {R.}~\bibnamefont {Cheng}}, \bibinfo {author}
  {\bibfnamefont {K.~L.}\ \bibnamefont {Seyler}}, \bibinfo {author}
  {\bibfnamefont {D.}~\bibnamefont {Zhong}}, \bibinfo {author} {\bibfnamefont
  {E.}~\bibnamefont {Schmidgall}}, \bibinfo {author} {\bibfnamefont {M.~A.}\
  \bibnamefont {McGuire}}, \bibinfo {author} {\bibfnamefont {D.~H.}\
  \bibnamefont {Cobden}}, \bibinfo {author} {\bibfnamefont {W.}~\bibnamefont
  {Yao}}, \bibinfo {author} {\bibfnamefont {D.}~\bibnamefont {Xiao}}, \bibinfo
  {author} {\bibfnamefont {P.}~\bibnamefont {Jarillo-Herrero}}, \ and\ \bibinfo
  {author} {\bibfnamefont {X.}~\bibnamefont {Xu}},\ }\href
  {http://dx.doi.org/10.1038/nature22391} {\bibfield  {journal} {\bibinfo
  {journal} {Nature}\ }\textbf {\bibinfo {volume} {546}},\ \bibinfo {pages}
  {270} (\bibinfo {year} {2017})}\BibitemShut {NoStop}%
\bibitem [{\citenamefont {Gong}\ \emph {et~al.}(2017)\citenamefont {Gong},
  \citenamefont {Li}, \citenamefont {Li}, \citenamefont {Ji}, \citenamefont
  {Stern}, \citenamefont {Xia}, \citenamefont {Cao}, \citenamefont {Bao},
  \citenamefont {Wang}, \citenamefont {Wang}, \citenamefont {Qiu},
  \citenamefont {Cava}, \citenamefont {Louie}, \citenamefont {Xia},\ and\
  \citenamefont {Zhang}}]{Gong2017}%
  \BibitemOpen
  \bibfield  {author} {\bibinfo {author} {\bibfnamefont {C.}~\bibnamefont
  {Gong}}, \bibinfo {author} {\bibfnamefont {L.}~\bibnamefont {Li}}, \bibinfo
  {author} {\bibfnamefont {Z.}~\bibnamefont {Li}}, \bibinfo {author}
  {\bibfnamefont {H.}~\bibnamefont {Ji}}, \bibinfo {author} {\bibfnamefont
  {A.}~\bibnamefont {Stern}}, \bibinfo {author} {\bibfnamefont
  {Y.}~\bibnamefont {Xia}}, \bibinfo {author} {\bibfnamefont {T.}~\bibnamefont
  {Cao}}, \bibinfo {author} {\bibfnamefont {W.}~\bibnamefont {Bao}}, \bibinfo
  {author} {\bibfnamefont {C.}~\bibnamefont {Wang}}, \bibinfo {author}
  {\bibfnamefont {Y.}~\bibnamefont {Wang}}, \bibinfo {author} {\bibfnamefont
  {Z.~Q.}\ \bibnamefont {Qiu}}, \bibinfo {author} {\bibfnamefont {R.~J.}\
  \bibnamefont {Cava}}, \bibinfo {author} {\bibfnamefont {S.~G.}\ \bibnamefont
  {Louie}}, \bibinfo {author} {\bibfnamefont {J.}~\bibnamefont {Xia}}, \ and\
  \bibinfo {author} {\bibfnamefont {X.}~\bibnamefont {Zhang}},\ }\href
  {http://dx.doi.org/10.1038/nature22060} {\bibfield  {journal} {\bibinfo
  {journal} {Nature}\ }\textbf {\bibinfo {volume} {546}},\ \bibinfo {pages}
  {265} (\bibinfo {year} {2017})}\BibitemShut {NoStop}%
\bibitem [{\citenamefont {Lee}\ \emph {et~al.}(2006)\citenamefont {Lee},
  \citenamefont {Nagaosa},\ and\ \citenamefont {Wen}}]{Lee2006}%
  \BibitemOpen
  \bibfield  {author} {\bibinfo {author} {\bibfnamefont {P.~A.}\ \bibnamefont
  {Lee}}, \bibinfo {author} {\bibfnamefont {N.}~\bibnamefont {Nagaosa}}, \ and\
  \bibinfo {author} {\bibfnamefont {X.-G.}\ \bibnamefont {Wen}},\ }\href
  {\doibase 10.1103/RevModPhys.78.17} {\bibfield  {journal} {\bibinfo
  {journal} {Rev. Mod. Phys.}\ }\textbf {\bibinfo {volume} {78}},\ \bibinfo
  {pages} {17} (\bibinfo {year} {2006})}\BibitemShut {NoStop}%
\bibitem [{\citenamefont {Stewart}(1984)}]{Stewart1984}%
  \BibitemOpen
  \bibfield  {author} {\bibinfo {author} {\bibfnamefont {G.~R.}\ \bibnamefont
  {Stewart}},\ }\href {\doibase 10.1103/RevModPhys.56.755} {\bibfield
  {journal} {\bibinfo  {journal} {Rev. Mod. Phys.}\ }\textbf {\bibinfo {volume}
  {56}},\ \bibinfo {pages} {755} (\bibinfo {year} {1984})}\BibitemShut
  {NoStop}%
\bibitem [{\citenamefont {Lee}(2008)}]{Lee2008}%
  \BibitemOpen
  \bibfield  {author} {\bibinfo {author} {\bibfnamefont {P.~A.}\ \bibnamefont
  {Lee}},\ }\href {\doibase 10.1088/0034-4885/71/1/012501} {\bibfield
  {journal} {\bibinfo  {journal} {Rep. Prog. Phys.}\ }\textbf {\bibinfo
  {volume} {71}},\ \bibinfo {pages} {012501} (\bibinfo {year}
  {2008})}\BibitemShut {NoStop}%
\bibitem [{\citenamefont {Cao}\ \emph {et~al.}(2018{\natexlab{a}})\citenamefont
  {Cao}, \citenamefont {Fatemi}, \citenamefont {Demir}, \citenamefont {Fang},
  \citenamefont {Tomarken}, \citenamefont {Luo}, \citenamefont
  {Sanchez-Yamagishi}, \citenamefont {Watanabe}, \citenamefont {Taniguchi},
  \citenamefont {Kaxiras}, \citenamefont {Ashoori},\ and\ \citenamefont
  {Jarillo-Herrero}}]{Cao2018}%
  \BibitemOpen
  \bibfield  {author} {\bibinfo {author} {\bibfnamefont {Y.}~\bibnamefont
  {Cao}}, \bibinfo {author} {\bibfnamefont {V.}~\bibnamefont {Fatemi}},
  \bibinfo {author} {\bibfnamefont {A.}~\bibnamefont {Demir}}, \bibinfo
  {author} {\bibfnamefont {S.}~\bibnamefont {Fang}}, \bibinfo {author}
  {\bibfnamefont {S.~L.}\ \bibnamefont {Tomarken}}, \bibinfo {author}
  {\bibfnamefont {J.~Y.}\ \bibnamefont {Luo}}, \bibinfo {author} {\bibfnamefont
  {J.~D.}\ \bibnamefont {Sanchez-Yamagishi}}, \bibinfo {author} {\bibfnamefont
  {K.}~\bibnamefont {Watanabe}}, \bibinfo {author} {\bibfnamefont
  {T.}~\bibnamefont {Taniguchi}}, \bibinfo {author} {\bibfnamefont
  {E.}~\bibnamefont {Kaxiras}}, \bibinfo {author} {\bibfnamefont {R.~C.}\
  \bibnamefont {Ashoori}}, \ and\ \bibinfo {author} {\bibfnamefont
  {P.}~\bibnamefont {Jarillo-Herrero}},\ }\href
  {http://dx.doi.org/10.1038/nature26154} {\bibfield  {journal} {\bibinfo
  {journal} {Nature}\ }\textbf {\bibinfo {volume} {556}},\ \bibinfo {pages}
  {80} (\bibinfo {year} {2018}{\natexlab{a}})}\BibitemShut {NoStop}%
\bibitem [{\citenamefont {Cao}\ \emph {et~al.}(2018{\natexlab{b}})\citenamefont
  {Cao}, \citenamefont {Fatemi}, \citenamefont {Fang}, \citenamefont
  {Watanabe}, \citenamefont {Taniguchi}, \citenamefont {Kaxiras},\ and\
  \citenamefont {Jarillo-Herrero}}]{Cao2018a}%
  \BibitemOpen
  \bibfield  {author} {\bibinfo {author} {\bibfnamefont {Y.}~\bibnamefont
  {Cao}}, \bibinfo {author} {\bibfnamefont {V.}~\bibnamefont {Fatemi}},
  \bibinfo {author} {\bibfnamefont {S.}~\bibnamefont {Fang}}, \bibinfo {author}
  {\bibfnamefont {K.}~\bibnamefont {Watanabe}}, \bibinfo {author}
  {\bibfnamefont {T.}~\bibnamefont {Taniguchi}}, \bibinfo {author}
  {\bibfnamefont {E.}~\bibnamefont {Kaxiras}}, \ and\ \bibinfo {author}
  {\bibfnamefont {P.}~\bibnamefont {Jarillo-Herrero}},\ }\href
  {http://dx.doi.org/10.1038/nature26160} {\bibfield  {journal} {\bibinfo
  {journal} {Nature}\ }\textbf {\bibinfo {volume} {556}},\ \bibinfo {pages}
  {43} (\bibinfo {year} {2018}{\natexlab{b}})}\BibitemShut {NoStop}%
\bibitem [{\citenamefont {Marseglia}(1983)}]{Marseglia1983}%
  \BibitemOpen
  \bibfield  {author} {\bibinfo {author} {\bibfnamefont {E.~A.}\ \bibnamefont
  {Marseglia}},\ }\href {\doibase 10.1080/01442358309353343} {\bibfield
  {journal} {\bibinfo  {journal} {Int. Rev. Phys. Chem.}\ }\textbf {\bibinfo
  {volume} {3}},\ \bibinfo {pages} {177} (\bibinfo {year} {1983})}\BibitemShut
  {NoStop}%
\bibitem [{\citenamefont {Friend}\ and\ \citenamefont
  {Yoffe}(1987)}]{Friend1987}%
  \BibitemOpen
  \bibfield  {author} {\bibinfo {author} {\bibfnamefont {R.}~\bibnamefont
  {Friend}}\ and\ \bibinfo {author} {\bibfnamefont {A.}~\bibnamefont {Yoffe}},\
  }\href {\doibase 10.1080/00018738700101951} {\bibfield  {journal} {\bibinfo
  {journal} {Adv. Phys.}\ }\textbf {\bibinfo {volume} {36}},\ \bibinfo {pages}
  {1} (\bibinfo {year} {1987})}\BibitemShut {NoStop}%
\bibitem [{\citenamefont {Dresselhaus}\ and\ \citenamefont
  {Dresselhaus}(2002)}]{Dresselhaus2002}%
  \BibitemOpen
  \bibfield  {author} {\bibinfo {author} {\bibfnamefont {M.~S.}\ \bibnamefont
  {Dresselhaus}}\ and\ \bibinfo {author} {\bibfnamefont {G.}~\bibnamefont
  {Dresselhaus}},\ }\href@noop {} {\bibfield  {journal} {\bibinfo  {journal}
  {Adv. Phys.}\ }\textbf {\bibinfo {volume} {51}},\ \bibinfo {pages} {1}
  (\bibinfo {year} {2002})}\BibitemShut {NoStop}%
\bibitem [{\citenamefont {Sugawara}\ \emph {et~al.}(2011)\citenamefont
  {Sugawara}, \citenamefont {Kanetani}, \citenamefont {Sato},\ and\
  \citenamefont {Takahashi}}]{Sugawara2011}%
  \BibitemOpen
  \bibfield  {author} {\bibinfo {author} {\bibfnamefont {K.}~\bibnamefont
  {Sugawara}}, \bibinfo {author} {\bibfnamefont {K.}~\bibnamefont {Kanetani}},
  \bibinfo {author} {\bibfnamefont {T.}~\bibnamefont {Sato}}, \ and\ \bibinfo
  {author} {\bibfnamefont {T.}~\bibnamefont {Takahashi}},\ }\href {\doibase
  10.1063/1.3582814} {\bibfield  {journal} {\bibinfo  {journal} {AIP Advances}\
  }\textbf {\bibinfo {volume} {1}},\ \bibinfo {pages} {022103} (\bibinfo {year}
  {2011})}\BibitemShut {NoStop}%
\bibitem [{\citenamefont {Kanetani}\ \emph {et~al.}(2012)\citenamefont
  {Kanetani}, \citenamefont {Sugawara}, \citenamefont {Sato}, \citenamefont
  {Shimizu}, \citenamefont {Iwaya}, \citenamefont {Hitosugi},\ and\
  \citenamefont {Takahashi}}]{Kanetani2012}%
  \BibitemOpen
  \bibfield  {author} {\bibinfo {author} {\bibfnamefont {K.}~\bibnamefont
  {Kanetani}}, \bibinfo {author} {\bibfnamefont {K.}~\bibnamefont {Sugawara}},
  \bibinfo {author} {\bibfnamefont {T.}~\bibnamefont {Sato}}, \bibinfo {author}
  {\bibfnamefont {R.}~\bibnamefont {Shimizu}}, \bibinfo {author} {\bibfnamefont
  {K.}~\bibnamefont {Iwaya}}, \bibinfo {author} {\bibfnamefont
  {T.}~\bibnamefont {Hitosugi}}, \ and\ \bibinfo {author} {\bibfnamefont
  {T.}~\bibnamefont {Takahashi}},\ }\href {\doibase 10.1073/pnas.1208889109}
  {\bibfield  {journal} {\bibinfo  {journal} {Proc. Natl. Acad. Sci.}\ }\textbf
  {\bibinfo {volume} {109}},\ \bibinfo {pages} {19610} (\bibinfo {year}
  {2012})}\BibitemShut {NoStop}%
\bibitem [{\citenamefont {Murphy}\ \emph {et~al.}(1977)\citenamefont {Murphy},
  \citenamefont {Cros}, \citenamefont {Salvo},\ and\ \citenamefont
  {Waszczak}}]{Murphy1977}%
  \BibitemOpen
  \bibfield  {author} {\bibinfo {author} {\bibfnamefont {D.~W.}\ \bibnamefont
  {Murphy}}, \bibinfo {author} {\bibfnamefont {C.}~\bibnamefont {Cros}},
  \bibinfo {author} {\bibfnamefont {F.~J.~D.}\ \bibnamefont {Salvo}}, \ and\
  \bibinfo {author} {\bibfnamefont {J.~V.}\ \bibnamefont {Waszczak}},\ }\href
  {\doibase 10.1021/ic50178a008} {\bibfield  {journal} {\bibinfo  {journal}
  {Inorg. Chem.}\ }\textbf {\bibinfo {volume} {16}},\ \bibinfo {pages} {3027}
  (\bibinfo {year} {1977})}\BibitemShut {NoStop}%
\bibitem [{\citenamefont {Niu}\ \emph {et~al.}(2017)\citenamefont {Niu},
  \citenamefont {Yan}, \citenamefont {Ji}, \citenamefont {Liu}, \citenamefont
  {Li}, \citenamefont {Gao}, \citenamefont {Zhang}, \citenamefont {Yu},\ and\
  \citenamefont {Wu}}]{Niu2017a}%
  \BibitemOpen
  \bibfield  {author} {\bibinfo {author} {\bibfnamefont {J.}~\bibnamefont
  {Niu}}, \bibinfo {author} {\bibfnamefont {B.}~\bibnamefont {Yan}}, \bibinfo
  {author} {\bibfnamefont {Q.}~\bibnamefont {Ji}}, \bibinfo {author}
  {\bibfnamefont {Z.}~\bibnamefont {Liu}}, \bibinfo {author} {\bibfnamefont
  {M.}~\bibnamefont {Li}}, \bibinfo {author} {\bibfnamefont {P.}~\bibnamefont
  {Gao}}, \bibinfo {author} {\bibfnamefont {Y.}~\bibnamefont {Zhang}}, \bibinfo
  {author} {\bibfnamefont {D.}~\bibnamefont {Yu}}, \ and\ \bibinfo {author}
  {\bibfnamefont {X.}~\bibnamefont {Wu}},\ }\href {\doibase
  10.1103/PhysRevB.96.075402} {\bibfield  {journal} {\bibinfo  {journal} {Phys.
  Rev. B}\ }\textbf {\bibinfo {volume} {96}},\ \bibinfo {pages} {075402}
  (\bibinfo {year} {2017})}\BibitemShut {NoStop}%
\bibitem [{\citenamefont {Jia}\ \emph {et~al.}(2016)\citenamefont {Jia},
  \citenamefont {Li}, \citenamefont {Li}, \citenamefont {Shi}, \citenamefont
  {Liao}, \citenamefont {Yu},\ and\ \citenamefont {Wu}}]{Jia2016}%
  \BibitemOpen
  \bibfield  {author} {\bibinfo {author} {\bibfnamefont {Z.}~\bibnamefont
  {Jia}}, \bibinfo {author} {\bibfnamefont {C.}~\bibnamefont {Li}}, \bibinfo
  {author} {\bibfnamefont {X.}~\bibnamefont {Li}}, \bibinfo {author}
  {\bibfnamefont {J.}~\bibnamefont {Shi}}, \bibinfo {author} {\bibfnamefont
  {Z.}~\bibnamefont {Liao}}, \bibinfo {author} {\bibfnamefont {D.}~\bibnamefont
  {Yu}}, \ and\ \bibinfo {author} {\bibfnamefont {X.}~\bibnamefont {Wu}},\
  }\href {http://dx.doi.org/10.1038/ncomms13013} {\bibfield  {journal}
  {\bibinfo  {journal} {Nat. Commun.}\ }\textbf {\bibinfo {volume} {7}},\
  \bibinfo {pages} {13013} (\bibinfo {year} {2016})}\BibitemShut {NoStop}%
\bibitem [{\citenamefont {Silbernagel}\ \emph {et~al.}(1975)\citenamefont
  {Silbernagel}, \citenamefont {Levy},\ and\ \citenamefont
  {Gamble}}]{Silbernagel1975}%
  \BibitemOpen
  \bibfield  {author} {\bibinfo {author} {\bibfnamefont {B.~G.}\ \bibnamefont
  {Silbernagel}}, \bibinfo {author} {\bibfnamefont {R.~B.}\ \bibnamefont
  {Levy}}, \ and\ \bibinfo {author} {\bibfnamefont {F.~R.}\ \bibnamefont
  {Gamble}},\ }\href {\doibase 10.1103/PhysRevB.11.4563} {\bibfield  {journal}
  {\bibinfo  {journal} {Phys. Rev. B}\ }\textbf {\bibinfo {volume} {11}},\
  \bibinfo {pages} {4563} (\bibinfo {year} {1975})}\BibitemShut {NoStop}%
\bibitem [{\citenamefont {Fujimori}\ \emph {et~al.}(1991)\citenamefont
  {Fujimori}, \citenamefont {Saeki},\ and\ \citenamefont
  {Nozaki}}]{Fujimori1991}%
  \BibitemOpen
  \bibfield  {author} {\bibinfo {author} {\bibfnamefont {A.}~\bibnamefont
  {Fujimori}}, \bibinfo {author} {\bibfnamefont {M.}~\bibnamefont {Saeki}}, \
  and\ \bibinfo {author} {\bibfnamefont {H.}~\bibnamefont {Nozaki}},\ }\href
  {\doibase 10.1103/PhysRevB.44.163} {\bibfield  {journal} {\bibinfo  {journal}
  {Phys. Rev. B}\ }\textbf {\bibinfo {volume} {44}},\ \bibinfo {pages} {163}
  (\bibinfo {year} {1991})}\BibitemShut {NoStop}%
\bibitem [{\citenamefont {Nozaki}\ \emph {et~al.}(1978)\citenamefont {Nozaki},
  \citenamefont {Umehara}, \citenamefont {Ishizawa}, \citenamefont {Saeki},
  \citenamefont {Mizoguchi},\ and\ \citenamefont {Nakahira}}]{Nozaki1978}%
  \BibitemOpen
  \bibfield  {author} {\bibinfo {author} {\bibfnamefont {H.}~\bibnamefont
  {Nozaki}}, \bibinfo {author} {\bibfnamefont {M.}~\bibnamefont {Umehara}},
  \bibinfo {author} {\bibfnamefont {Y.}~\bibnamefont {Ishizawa}}, \bibinfo
  {author} {\bibfnamefont {M.}~\bibnamefont {Saeki}}, \bibinfo {author}
  {\bibfnamefont {T.}~\bibnamefont {Mizoguchi}}, \ and\ \bibinfo {author}
  {\bibfnamefont {M.}~\bibnamefont {Nakahira}},\ }\href
  {http://www.sciencedirect.com/science/article/pii/0022369778901440}
  {\bibfield  {journal} {\bibinfo  {journal} {J. Phys. Chem. Solids}\ }\textbf
  {\bibinfo {volume} {39}},\ \bibinfo {pages} {851} (\bibinfo {year}
  {1978})}\BibitemShut {NoStop}%
\bibitem [{\citenamefont {Vries}\ and\ \citenamefont {Haas}(1973)}]{Vries1973}%
  \BibitemOpen
  \bibfield  {author} {\bibinfo {author} {\bibfnamefont {A.~D.}\ \bibnamefont
  {Vries}}\ and\ \bibinfo {author} {\bibfnamefont {C.}~\bibnamefont {Haas}},\
  }\href {\doibase http://dx.doi.org/10.1016/S0022-3697(73)80171-4} {\bibfield
  {journal} {\bibinfo  {journal} {J. Phys. Chem. Solids}\ }\textbf {\bibinfo
  {volume} {34}},\ \bibinfo {pages} {651} (\bibinfo {year} {1973})}\BibitemShut
  {NoStop}%
\bibitem [{\citenamefont {Nishihara}\ \emph {et~al.}(1977)\citenamefont
  {Nishihara}, \citenamefont {Yasuoka}, \citenamefont {Oka}, \citenamefont
  {Kosuge},\ and\ \citenamefont {Kachi}}]{Nishihara1977}%
  \BibitemOpen
  \bibfield  {author} {\bibinfo {author} {\bibfnamefont {H.}~\bibnamefont
  {Nishihara}}, \bibinfo {author} {\bibfnamefont {H.}~\bibnamefont {Yasuoka}},
  \bibinfo {author} {\bibfnamefont {Y.}~\bibnamefont {Oka}}, \bibinfo {author}
  {\bibfnamefont {K.}~\bibnamefont {Kosuge}}, \ and\ \bibinfo {author}
  {\bibfnamefont {S.}~\bibnamefont {Kachi}},\ }\href {\doibase
  10.1143/JPSJ.42.787} {\bibfield  {journal} {\bibinfo  {journal} {J. Phys.
  Soc. Jpn.}\ }\textbf {\bibinfo {volume} {42}},\ \bibinfo {pages} {787}
  (\bibinfo {year} {1977})}\BibitemShut {NoStop}%
\bibitem [{\citenamefont {Kitaoka}\ and\ \citenamefont
  {Yasuoka}(1980)}]{Kitaoka1980}%
  \BibitemOpen
  \bibfield  {author} {\bibinfo {author} {\bibfnamefont {Y.}~\bibnamefont
  {Kitaoka}}\ and\ \bibinfo {author} {\bibfnamefont {H.}~\bibnamefont
  {Yasuoka}},\ }\href {\doibase 10.1143/JPSJ.48.1949} {\bibfield  {journal}
  {\bibinfo  {journal} {J. Phys. Soc. Jpn.}\ }\textbf {\bibinfo {volume}
  {48}},\ \bibinfo {pages} {1949} (\bibinfo {year} {1980})}\BibitemShut
  {NoStop}%
\bibitem [{\citenamefont {Katsuta}\ \emph {et~al.}(1979)\citenamefont
  {Katsuta}, \citenamefont {McLellan},\ and\ \citenamefont
  {Suzuki}}]{Katsuta1979}%
  \BibitemOpen
  \bibfield  {author} {\bibinfo {author} {\bibfnamefont {H.}~\bibnamefont
  {Katsuta}}, \bibinfo {author} {\bibfnamefont {R.~B.}\ \bibnamefont
  {McLellan}}, \ and\ \bibinfo {author} {\bibfnamefont {K.}~\bibnamefont
  {Suzuki}},\ }\href {\doibase http://dx.doi.org/10.1016/0022-3697(79)90143-4}
  {\bibfield  {journal} {\bibinfo  {journal} {J. Phys. Chem. Solids}\ }\textbf
  {\bibinfo {volume} {40}},\ \bibinfo {pages} {1089} (\bibinfo {year}
  {1979})}\BibitemShut {NoStop}%
\bibitem [{\citenamefont {Funahashi}\ \emph {et~al.}(1981)\citenamefont
  {Funahashi}, \citenamefont {Nozaki},\ and\ \citenamefont
  {Kawada}}]{Funahashi1981}%
  \BibitemOpen
  \bibfield  {author} {\bibinfo {author} {\bibfnamefont {S.}~\bibnamefont
  {Funahashi}}, \bibinfo {author} {\bibfnamefont {H.}~\bibnamefont {Nozaki}}, \
  and\ \bibinfo {author} {\bibfnamefont {I.}~\bibnamefont {Kawada}},\ }\href
  {http://www.sciencedirect.com/science/article/pii/0022369781900640}
  {\bibfield  {journal} {\bibinfo  {journal} {J. Phys. Chem. Solids}\ }\textbf
  {\bibinfo {volume} {42}},\ \bibinfo {pages} {1009} (\bibinfo {year}
  {1981})}\BibitemShut {NoStop}%
\bibitem [{\citenamefont {Nakanishi}\ \emph {et~al.}(2000)\citenamefont
  {Nakanishi}, \citenamefont {Yoshimura}, \citenamefont {Kosuge}, \citenamefont
  {Goto}, \citenamefont {Fujii},\ and\ \citenamefont {Takada}}]{Nakanishi2000}%
  \BibitemOpen
  \bibfield  {author} {\bibinfo {author} {\bibfnamefont {M.}~\bibnamefont
  {Nakanishi}}, \bibinfo {author} {\bibfnamefont {K.}~\bibnamefont
  {Yoshimura}}, \bibinfo {author} {\bibfnamefont {K.}~\bibnamefont {Kosuge}},
  \bibinfo {author} {\bibfnamefont {T.}~\bibnamefont {Goto}}, \bibinfo {author}
  {\bibfnamefont {T.}~\bibnamefont {Fujii}}, \ and\ \bibinfo {author}
  {\bibfnamefont {J.}~\bibnamefont {Takada}},\ }\href {\doibase
  http://dx.doi.org/10.1016/S0304-8853(00)00509-6} {\bibfield  {journal}
  {\bibinfo  {journal} {J. Magn. Magn. Mater.}\ }\textbf {\bibinfo {volume}
  {221}},\ \bibinfo {pages} {301} (\bibinfo {year} {2000})}\BibitemShut
  {NoStop}%
\bibitem [{\citenamefont {Barua}\ \emph {et~al.}(2017)\citenamefont {Barua},
  \citenamefont {Hatnean}, \citenamefont {Lees},\ and\ \citenamefont
  {Balakrishnan}}]{Barua2017}%
  \BibitemOpen
  \bibfield  {author} {\bibinfo {author} {\bibfnamefont {S.}~\bibnamefont
  {Barua}}, \bibinfo {author} {\bibfnamefont {M.~C.}\ \bibnamefont {Hatnean}},
  \bibinfo {author} {\bibfnamefont {M.~R.}\ \bibnamefont {Lees}}, \ and\
  \bibinfo {author} {\bibfnamefont {G.}~\bibnamefont {Balakrishnan}},\ }\href
  {https://doi.org/10.1038/s41598-017-11247-4} {\bibfield  {journal} {\bibinfo
  {journal} {Sci. Rep.}\ }\textbf {\bibinfo {volume} {7}},\ \bibinfo {pages}
  {10964} (\bibinfo {year} {2017})}\BibitemShut {NoStop}%
\bibitem [{\citenamefont {Hossain}\ \emph {et~al.}(1999)\citenamefont
  {Hossain}, \citenamefont {Ohmoto}, \citenamefont {Umeo}, \citenamefont {Iga},
  \citenamefont {Suzuki}, \citenamefont {Takabatake}, \citenamefont
  {Takamoto},\ and\ \citenamefont {Kindo}}]{Hossain1999}%
  \BibitemOpen
  \bibfield  {author} {\bibinfo {author} {\bibfnamefont {Z.}~\bibnamefont
  {Hossain}}, \bibinfo {author} {\bibfnamefont {H.}~\bibnamefont {Ohmoto}},
  \bibinfo {author} {\bibfnamefont {K.}~\bibnamefont {Umeo}}, \bibinfo {author}
  {\bibfnamefont {F.}~\bibnamefont {Iga}}, \bibinfo {author} {\bibfnamefont
  {T.}~\bibnamefont {Suzuki}}, \bibinfo {author} {\bibfnamefont
  {T.}~\bibnamefont {Takabatake}}, \bibinfo {author} {\bibfnamefont
  {N.}~\bibnamefont {Takamoto}}, \ and\ \bibinfo {author} {\bibfnamefont
  {K.}~\bibnamefont {Kindo}},\ }\href {\doibase 10.1103/PhysRevB.60.10383}
  {\bibfield  {journal} {\bibinfo  {journal} {Phys. Rev. B}\ }\textbf {\bibinfo
  {volume} {60}},\ \bibinfo {pages} {10383} (\bibinfo {year}
  {1999})}\BibitemShut {NoStop}%
\bibitem [{\citenamefont {Fisk}\ \emph {et~al.}(1986)\citenamefont {Fisk},
  \citenamefont {Ott}, \citenamefont {Rice},\ and\ \citenamefont
  {Smith}}]{Fisk1986}%
  \BibitemOpen
  \bibfield  {author} {\bibinfo {author} {\bibfnamefont {Z.}~\bibnamefont
  {Fisk}}, \bibinfo {author} {\bibfnamefont {H.~R.}\ \bibnamefont {Ott}},
  \bibinfo {author} {\bibfnamefont {T.~M.}\ \bibnamefont {Rice}}, \ and\
  \bibinfo {author} {\bibfnamefont {J.~L.}\ \bibnamefont {Smith}},\ }\href
  {http://dx.doi.org/10.1038/320124a0} {\bibfield  {journal} {\bibinfo
  {journal} {Nature}\ }\textbf {\bibinfo {volume} {320}},\ \bibinfo {pages}
  {124} (\bibinfo {year} {1986})}\BibitemShut {NoStop}%
\bibitem [{\citenamefont {Kondo}\ \emph {et~al.}(1997)\citenamefont {Kondo},
  \citenamefont {Johnston}, \citenamefont {Swenson}, \citenamefont {Borsa},
  \citenamefont {Mahajan}, \citenamefont {Miller}, \citenamefont {Gu},
  \citenamefont {Goldman}, \citenamefont {Maple}, \citenamefont {Gajewski},
  \citenamefont {Freeman}, \citenamefont {Dilley}, \citenamefont {Dickey},
  \citenamefont {Merrin}, \citenamefont {Kojima}, \citenamefont {Luke},
  \citenamefont {Uemura}, \citenamefont {Chmaissem},\ and\ \citenamefont
  {Jorgensen}}]{Kondo1997}%
  \BibitemOpen
  \bibfield  {author} {\bibinfo {author} {\bibfnamefont {S.}~\bibnamefont
  {Kondo}}, \bibinfo {author} {\bibfnamefont {D.~C.}\ \bibnamefont {Johnston}},
  \bibinfo {author} {\bibfnamefont {C.~A.}\ \bibnamefont {Swenson}}, \bibinfo
  {author} {\bibfnamefont {F.}~\bibnamefont {Borsa}}, \bibinfo {author}
  {\bibfnamefont {A.~V.}\ \bibnamefont {Mahajan}}, \bibinfo {author}
  {\bibfnamefont {L.~L.}\ \bibnamefont {Miller}}, \bibinfo {author}
  {\bibfnamefont {T.}~\bibnamefont {Gu}}, \bibinfo {author} {\bibfnamefont
  {A.~I.}\ \bibnamefont {Goldman}}, \bibinfo {author} {\bibfnamefont {M.~B.}\
  \bibnamefont {Maple}}, \bibinfo {author} {\bibfnamefont {D.~A.}\ \bibnamefont
  {Gajewski}}, \bibinfo {author} {\bibfnamefont {E.~J.}\ \bibnamefont
  {Freeman}}, \bibinfo {author} {\bibfnamefont {N.~R.}\ \bibnamefont {Dilley}},
  \bibinfo {author} {\bibfnamefont {R.~P.}\ \bibnamefont {Dickey}}, \bibinfo
  {author} {\bibfnamefont {J.}~\bibnamefont {Merrin}}, \bibinfo {author}
  {\bibfnamefont {K.}~\bibnamefont {Kojima}}, \bibinfo {author} {\bibfnamefont
  {G.~M.}\ \bibnamefont {Luke}}, \bibinfo {author} {\bibfnamefont {Y.~J.}\
  \bibnamefont {Uemura}}, \bibinfo {author} {\bibfnamefont {O.}~\bibnamefont
  {Chmaissem}}, \ and\ \bibinfo {author} {\bibfnamefont {J.~D.}\ \bibnamefont
  {Jorgensen}},\ }\href {\doibase 10.1103/PhysRevLett.78.3729} {\bibfield
  {journal} {\bibinfo  {journal} {Phys. Rev. Lett.}\ }\textbf {\bibinfo
  {volume} {78}},\ \bibinfo {pages} {3729} (\bibinfo {year}
  {1997})}\BibitemShut {NoStop}%
\bibitem [{\citenamefont {Pikul}\ \emph {et~al.}(2003)\citenamefont {Pikul},
  \citenamefont {Kaczorowski}, \citenamefont {Plackowski}, \citenamefont
  {Czopnik}, \citenamefont {Michor}, \citenamefont {Bauer}, \citenamefont
  {Hilscher}, \citenamefont {Rogl},\ and\ \citenamefont {Grin}}]{Pikul2003}%
  \BibitemOpen
  \bibfield  {author} {\bibinfo {author} {\bibfnamefont {A.~P.}\ \bibnamefont
  {Pikul}}, \bibinfo {author} {\bibfnamefont {D.}~\bibnamefont {Kaczorowski}},
  \bibinfo {author} {\bibfnamefont {T.}~\bibnamefont {Plackowski}}, \bibinfo
  {author} {\bibfnamefont {A.}~\bibnamefont {Czopnik}}, \bibinfo {author}
  {\bibfnamefont {H.}~\bibnamefont {Michor}}, \bibinfo {author} {\bibfnamefont
  {E.}~\bibnamefont {Bauer}}, \bibinfo {author} {\bibfnamefont
  {G.}~\bibnamefont {Hilscher}}, \bibinfo {author} {\bibfnamefont
  {P.}~\bibnamefont {Rogl}}, \ and\ \bibinfo {author} {\bibfnamefont
  {Y.}~\bibnamefont {Grin}},\ }\href {\doibase 10.1103/PhysRevB.67.224417}
  {\bibfield  {journal} {\bibinfo  {journal} {Phys. Rev. B}\ }\textbf {\bibinfo
  {volume} {67}},\ \bibinfo {pages} {224417} (\bibinfo {year}
  {2003})}\BibitemShut {NoStop}%
\bibitem [{\citenamefont {Zhou}\ \emph {et~al.}(2018)\citenamefont {Zhou},
  \citenamefont {Xu}, \citenamefont {Li}, \citenamefont {Sankar}, \citenamefont
  {Zhang}, \citenamefont {Qian}, \citenamefont {Cao}, \citenamefont {Dai},
  \citenamefont {Lu},\ and\ \citenamefont {Xu}}]{Zhou2018}%
  \BibitemOpen
  \bibfield  {author} {\bibinfo {author} {\bibfnamefont {W.}~\bibnamefont
  {Zhou}}, \bibinfo {author} {\bibfnamefont {C.~Q.}\ \bibnamefont {Xu}},
  \bibinfo {author} {\bibfnamefont {B.}~\bibnamefont {Li}}, \bibinfo {author}
  {\bibfnamefont {R.}~\bibnamefont {Sankar}}, \bibinfo {author} {\bibfnamefont
  {F.~M.}\ \bibnamefont {Zhang}}, \bibinfo {author} {\bibfnamefont
  {B.}~\bibnamefont {Qian}}, \bibinfo {author} {\bibfnamefont {C.}~\bibnamefont
  {Cao}}, \bibinfo {author} {\bibfnamefont {J.~H.}\ \bibnamefont {Dai}},
  \bibinfo {author} {\bibfnamefont {J.}~\bibnamefont {Lu}}, \ and\ \bibinfo
  {author} {\bibfnamefont {X.}~\bibnamefont {Xu}},\ }\href@noop {} {\bibfield
  {journal} {\bibinfo  {journal} {ArXiv e-prints}\ ,\ \bibinfo {pages}
  {1802.05060}} (\bibinfo {year} {2018})}\BibitemShut {NoStop}%
\bibitem [{\citenamefont {Continentino}\ \emph {et~al.}(2001)\citenamefont
  {Continentino}, \citenamefont {Medeiros}, \citenamefont {Orlando},
  \citenamefont {Fontes},\ and\ \citenamefont
  {Baggio-Saitovitch}}]{Continentino2001}%
  \BibitemOpen
  \bibfield  {author} {\bibinfo {author} {\bibfnamefont {M.~A.}\ \bibnamefont
  {Continentino}}, \bibinfo {author} {\bibfnamefont {S.~N.~d.}\ \bibnamefont
  {Medeiros}}, \bibinfo {author} {\bibfnamefont {M.~T.~D.}\ \bibnamefont
  {Orlando}}, \bibinfo {author} {\bibfnamefont {M.~B.}\ \bibnamefont {Fontes}},
  \ and\ \bibinfo {author} {\bibfnamefont {E.~M.}\ \bibnamefont
  {Baggio-Saitovitch}},\ }\href {\doibase 10.1103/PhysRevB.64.012404}
  {\bibfield  {journal} {\bibinfo  {journal} {Phys. Rev. B}\ }\textbf {\bibinfo
  {volume} {64}},\ \bibinfo {pages} {012404} (\bibinfo {year}
  {2001})}\BibitemShut {NoStop}%
\bibitem [{\citenamefont {Szlawska}\ and\ \citenamefont
  {Kaczorowski}(2012)}]{Szlawska2012}%
  \BibitemOpen
  \bibfield  {author} {\bibinfo {author} {\bibfnamefont {M.}~\bibnamefont
  {Szlawska}}\ and\ \bibinfo {author} {\bibfnamefont {D.}~\bibnamefont
  {Kaczorowski}},\ }\href {\doibase 10.1103/PhysRevB.85.134423} {\bibfield
  {journal} {\bibinfo  {journal} {Phys. Rev. B}\ }\textbf {\bibinfo {volume}
  {85}},\ \bibinfo {pages} {134423} (\bibinfo {year} {2012})}\BibitemShut
  {NoStop}%
\bibitem [{\citenamefont {Li}\ \emph {et~al.}(2004)\citenamefont {Li},
  \citenamefont {Taillefer}, \citenamefont {Hawthorn}, \citenamefont {Tanatar},
  \citenamefont {Paglione}, \citenamefont {Sutherland}, \citenamefont {Hill},
  \citenamefont {Wang},\ and\ \citenamefont {Chen}}]{Li2004}%
  \BibitemOpen
  \bibfield  {author} {\bibinfo {author} {\bibfnamefont {S.~Y.}\ \bibnamefont
  {Li}}, \bibinfo {author} {\bibfnamefont {L.}~\bibnamefont {Taillefer}},
  \bibinfo {author} {\bibfnamefont {D.~G.}\ \bibnamefont {Hawthorn}}, \bibinfo
  {author} {\bibfnamefont {M.~A.}\ \bibnamefont {Tanatar}}, \bibinfo {author}
  {\bibfnamefont {J.}~\bibnamefont {Paglione}}, \bibinfo {author}
  {\bibfnamefont {M.}~\bibnamefont {Sutherland}}, \bibinfo {author}
  {\bibfnamefont {R.~W.}\ \bibnamefont {Hill}}, \bibinfo {author}
  {\bibfnamefont {C.~H.}\ \bibnamefont {Wang}}, \ and\ \bibinfo {author}
  {\bibfnamefont {X.~H.}\ \bibnamefont {Chen}},\ }\href {\doibase
  10.1103/PhysRevLett.93.056401} {\bibfield  {journal} {\bibinfo  {journal}
  {Phys. Rev. Lett.}\ }\textbf {\bibinfo {volume} {93}},\ \bibinfo {pages}
  {056401} (\bibinfo {year} {2004})}\BibitemShut {NoStop}%
\bibitem [{\citenamefont {Doniach}(1977)}]{Doniach1977}%
  \BibitemOpen
  \bibfield  {author} {\bibinfo {author} {\bibfnamefont {S.}~\bibnamefont
  {Doniach}},\ }\href
  {http://www.sciencedirect.com/science/article/pii/0378436377901905}
  {\bibfield  {journal} {\bibinfo  {journal} {Physica B+C}\ }\textbf {\bibinfo
  {volume} {91}},\ \bibinfo {pages} {231} (\bibinfo {year} {1977})}\BibitemShut
  {NoStop}%
\bibitem [{\citenamefont {Gauzzi}\ \emph {et~al.}(2014)\citenamefont {Gauzzi},
  \citenamefont {Sellam}, \citenamefont {Rousse}, \citenamefont {Klein},
  \citenamefont {Taverna}, \citenamefont {Giura}, \citenamefont {Calandra},
  \citenamefont {Loupias}, \citenamefont {Gozzo}, \citenamefont {Gilioli},
  \citenamefont {Bolzoni}, \citenamefont {Allodi}, \citenamefont {De~Renzi},
  \citenamefont {Calestani},\ and\ \citenamefont {Roy}}]{Gauzzi2014}%
  \BibitemOpen
  \bibfield  {author} {\bibinfo {author} {\bibfnamefont {A.}~\bibnamefont
  {Gauzzi}}, \bibinfo {author} {\bibfnamefont {A.}~\bibnamefont {Sellam}},
  \bibinfo {author} {\bibfnamefont {G.}~\bibnamefont {Rousse}}, \bibinfo
  {author} {\bibfnamefont {Y.}~\bibnamefont {Klein}}, \bibinfo {author}
  {\bibfnamefont {D.}~\bibnamefont {Taverna}}, \bibinfo {author} {\bibfnamefont
  {P.}~\bibnamefont {Giura}}, \bibinfo {author} {\bibfnamefont
  {M.}~\bibnamefont {Calandra}}, \bibinfo {author} {\bibfnamefont
  {G.}~\bibnamefont {Loupias}}, \bibinfo {author} {\bibfnamefont
  {F.}~\bibnamefont {Gozzo}}, \bibinfo {author} {\bibfnamefont
  {E.}~\bibnamefont {Gilioli}}, \bibinfo {author} {\bibfnamefont
  {F.}~\bibnamefont {Bolzoni}}, \bibinfo {author} {\bibfnamefont
  {G.}~\bibnamefont {Allodi}}, \bibinfo {author} {\bibfnamefont
  {R.}~\bibnamefont {De~Renzi}}, \bibinfo {author} {\bibfnamefont {G.~L.}\
  \bibnamefont {Calestani}}, \ and\ \bibinfo {author} {\bibfnamefont
  {P.}~\bibnamefont {Roy}},\ }\href {\doibase 10.1103/PhysRevB.89.235125}
  {\bibfield  {journal} {\bibinfo  {journal} {Phys. Rev. B}\ }\textbf {\bibinfo
  {volume} {89}},\ \bibinfo {pages} {235125} (\bibinfo {year}
  {2014})}\BibitemShut {NoStop}%
\bibitem [{\citenamefont {Kohn}\ and\ \citenamefont {Sham}(1965)}]{Kohn1965}%
  \BibitemOpen
  \bibfield  {author} {\bibinfo {author} {\bibfnamefont {W.}~\bibnamefont
  {Kohn}}\ and\ \bibinfo {author} {\bibfnamefont {L.~J.}\ \bibnamefont
  {Sham}},\ }\href {\doibase 10.1103/PhysRev.140.A1133} {\bibfield  {journal}
  {\bibinfo  {journal} {Phys. Rev.}\ }\textbf {\bibinfo {volume} {140}},\
  \bibinfo {pages} {A1133} (\bibinfo {year} {1965})}\BibitemShut {NoStop}%
\bibitem [{\citenamefont {Beck}\ and\ \citenamefont {Claus}(1970)}]{BECK1970}%
  \BibitemOpen
  \bibfield  {author} {\bibinfo {author} {\bibfnamefont {P.~A.}\ \bibnamefont
  {Beck}}\ and\ \bibinfo {author} {\bibfnamefont {H.}~\bibnamefont {Claus}},\
  }\href {\doibase 10.6028/jres.074A.035} {\bibfield  {journal} {\bibinfo
  {journal} {J. Res. Nat. Bur. Stand.}\ }\textbf {\bibinfo {volume} {A 74}},\
  \bibinfo {pages} {449} (\bibinfo {year} {1970})}\BibitemShut {NoStop}%
\bibitem [{\citenamefont {Savrasov}\ and\ \citenamefont
  {Savrasov}(1996)}]{Savrasov1996}%
  \BibitemOpen
  \bibfield  {author} {\bibinfo {author} {\bibfnamefont {S.~Y.}\ \bibnamefont
  {Savrasov}}\ and\ \bibinfo {author} {\bibfnamefont {D.~Y.}\ \bibnamefont
  {Savrasov}},\ }\href {\doibase 10.1103/PhysRevB.54.16487} {\bibfield
  {journal} {\bibinfo  {journal} {Phys. Rev. B}\ }\textbf {\bibinfo {volume}
  {54}},\ \bibinfo {pages} {16487} (\bibinfo {year} {1996})}\BibitemShut
  {NoStop}%
\bibitem [{\citenamefont {Hossain}\ \emph {et~al.}(2000)\citenamefont
  {Hossain}, \citenamefont {Hamashima}, \citenamefont {Umeo}, \citenamefont
  {Takabatake}, \citenamefont {Geibel},\ and\ \citenamefont
  {Steglich}}]{Hossain2000}%
  \BibitemOpen
  \bibfield  {author} {\bibinfo {author} {\bibfnamefont {Z.}~\bibnamefont
  {Hossain}}, \bibinfo {author} {\bibfnamefont {S.}~\bibnamefont {Hamashima}},
  \bibinfo {author} {\bibfnamefont {K.}~\bibnamefont {Umeo}}, \bibinfo {author}
  {\bibfnamefont {T.}~\bibnamefont {Takabatake}}, \bibinfo {author}
  {\bibfnamefont {C.}~\bibnamefont {Geibel}}, \ and\ \bibinfo {author}
  {\bibfnamefont {F.}~\bibnamefont {Steglich}},\ }\href {\doibase
  10.1103/PhysRevB.62.8950} {\bibfield  {journal} {\bibinfo  {journal} {Phys.
  Rev. B}\ }\textbf {\bibinfo {volume} {62}},\ \bibinfo {pages} {8950}
  (\bibinfo {year} {2000})}\BibitemShut {NoStop}%
\bibitem [{\citenamefont {Raquet}\ \emph {et~al.}(2002)\citenamefont {Raquet},
  \citenamefont {Viret}, \citenamefont {Sondergard}, \citenamefont {Cespedes},\
  and\ \citenamefont {Mamy}}]{Raquet2002}%
  \BibitemOpen
  \bibfield  {author} {\bibinfo {author} {\bibfnamefont {B.}~\bibnamefont
  {Raquet}}, \bibinfo {author} {\bibfnamefont {M.}~\bibnamefont {Viret}},
  \bibinfo {author} {\bibfnamefont {E.}~\bibnamefont {Sondergard}}, \bibinfo
  {author} {\bibfnamefont {O.}~\bibnamefont {Cespedes}}, \ and\ \bibinfo
  {author} {\bibfnamefont {R.}~\bibnamefont {Mamy}},\ }\href {\doibase
  10.1103/PhysRevB.66.024433} {\bibfield  {journal} {\bibinfo  {journal} {Phys.
  Rev. B}\ }\textbf {\bibinfo {volume} {66}},\ \bibinfo {pages} {024433}
  (\bibinfo {year} {2002})}\BibitemShut {NoStop}%
\bibitem [{\citenamefont {Madduri}\ and\ \citenamefont
  {Kaul}(2017)}]{Madduri2017}%
  \BibitemOpen
  \bibfield  {author} {\bibinfo {author} {\bibfnamefont {P.~V.~P.}\
  \bibnamefont {Madduri}}\ and\ \bibinfo {author} {\bibfnamefont {S.~N.}\
  \bibnamefont {Kaul}},\ }\href {\doibase 10.1103/PhysRevB.95.184402}
  {\bibfield  {journal} {\bibinfo  {journal} {Phys. Rev. B}\ }\textbf {\bibinfo
  {volume} {95}},\ \bibinfo {pages} {184402} (\bibinfo {year}
  {2017})}\BibitemShut {NoStop}%
\bibitem [{\citenamefont {Mentink}\ \emph {et~al.}(1996)\citenamefont
  {Mentink}, \citenamefont {Mason}, \citenamefont {S\"ullow}, \citenamefont
  {Nieuwenhuys}, \citenamefont {Menovsky}, \citenamefont {Mydosh},\ and\
  \citenamefont {Perenboom}}]{Mentink1996}%
  \BibitemOpen
  \bibfield  {author} {\bibinfo {author} {\bibfnamefont {S.~A.~M.}\
  \bibnamefont {Mentink}}, \bibinfo {author} {\bibfnamefont {T.~E.}\
  \bibnamefont {Mason}}, \bibinfo {author} {\bibfnamefont {S.}~\bibnamefont
  {S\"ullow}}, \bibinfo {author} {\bibfnamefont {G.~J.}\ \bibnamefont
  {Nieuwenhuys}}, \bibinfo {author} {\bibfnamefont {A.~A.}\ \bibnamefont
  {Menovsky}}, \bibinfo {author} {\bibfnamefont {J.~A.}\ \bibnamefont
  {Mydosh}}, \ and\ \bibinfo {author} {\bibfnamefont {J.~A. A.~J.}\
  \bibnamefont {Perenboom}},\ }\href {\doibase 10.1103/PhysRevB.53.R6014}
  {\bibfield  {journal} {\bibinfo  {journal} {Phys. Rev. B}\ }\textbf {\bibinfo
  {volume} {53}},\ \bibinfo {pages} {R6014} (\bibinfo {year}
  {1996})}\BibitemShut {NoStop}%
\bibitem [{\citenamefont {Jobiliong}\ \emph {et~al.}(2005)\citenamefont
  {Jobiliong}, \citenamefont {Brooks}, \citenamefont {Choi}, \citenamefont
  {Lee},\ and\ \citenamefont {Fisk}}]{Jobiliong2005}%
  \BibitemOpen
  \bibfield  {author} {\bibinfo {author} {\bibfnamefont {E.}~\bibnamefont
  {Jobiliong}}, \bibinfo {author} {\bibfnamefont {J.~S.}\ \bibnamefont
  {Brooks}}, \bibinfo {author} {\bibfnamefont {E.~S.}\ \bibnamefont {Choi}},
  \bibinfo {author} {\bibfnamefont {H.}~\bibnamefont {Lee}}, \ and\ \bibinfo
  {author} {\bibfnamefont {Z.}~\bibnamefont {Fisk}},\ }\href {\doibase
  10.1103/PhysRevB.72.104428} {\bibfield  {journal} {\bibinfo  {journal} {Phys.
  Rev. B}\ }\textbf {\bibinfo {volume} {72}},\ \bibinfo {pages} {104428}
  (\bibinfo {year} {2005})}\BibitemShut {NoStop}%
\bibitem [{\citenamefont {Kadowaki}\ and\ \citenamefont
  {Woods}(1986)}]{Kadowaki1986}%
  \BibitemOpen
  \bibfield  {author} {\bibinfo {author} {\bibfnamefont {K.}~\bibnamefont
  {Kadowaki}}\ and\ \bibinfo {author} {\bibfnamefont {S.}~\bibnamefont
  {Woods}},\ }\href {\doibase https://doi.org/10.1016/0038-1098(86)90785-4}
  {\bibfield  {journal} {\bibinfo  {journal} {Solid State Commun.}\ }\textbf
  {\bibinfo {volume} {58}},\ \bibinfo {pages} {507 } (\bibinfo {year}
  {1986})}\BibitemShut {NoStop}%
\bibitem [{\citenamefont {McWhan}\ \emph {et~al.}(1973)\citenamefont {McWhan},
  \citenamefont {Remeika}, \citenamefont {Bader}, \citenamefont {Triplett},\
  and\ \citenamefont {Phillips}}]{McWhan1973}%
  \BibitemOpen
  \bibfield  {author} {\bibinfo {author} {\bibfnamefont {D.~B.}\ \bibnamefont
  {McWhan}}, \bibinfo {author} {\bibfnamefont {J.~P.}\ \bibnamefont {Remeika}},
  \bibinfo {author} {\bibfnamefont {S.~D.}\ \bibnamefont {Bader}}, \bibinfo
  {author} {\bibfnamefont {B.~B.}\ \bibnamefont {Triplett}}, \ and\ \bibinfo
  {author} {\bibfnamefont {N.~E.}\ \bibnamefont {Phillips}},\ }\href {\doibase
  10.1103/PhysRevB.7.3079} {\bibfield  {journal} {\bibinfo  {journal} {Phys.
  Rev. B}\ }\textbf {\bibinfo {volume} {7}},\ \bibinfo {pages} {3079} (\bibinfo
  {year} {1973})}\BibitemShut {NoStop}%
\bibitem [{\citenamefont {Maeno}\ \emph {et~al.}(1997)\citenamefont {Maeno},
  \citenamefont {Yoshida}, \citenamefont {Hashimoto}, \citenamefont
  {Nishizaki}, \citenamefont {ichi Ikeda}, \citenamefont {Nohara},
  \citenamefont {Fujita}, \citenamefont {Mackenzie}, \citenamefont {Hussey},
  \citenamefont {Bednorz},\ and\ \citenamefont {Lichtenberg}}]{Maeno1997}%
  \BibitemOpen
  \bibfield  {author} {\bibinfo {author} {\bibfnamefont {Y.}~\bibnamefont
  {Maeno}}, \bibinfo {author} {\bibfnamefont {K.}~\bibnamefont {Yoshida}},
  \bibinfo {author} {\bibfnamefont {H.}~\bibnamefont {Hashimoto}}, \bibinfo
  {author} {\bibfnamefont {S.}~\bibnamefont {Nishizaki}}, \bibinfo {author}
  {\bibfnamefont {S.}~\bibnamefont {ichi Ikeda}}, \bibinfo {author}
  {\bibfnamefont {M.}~\bibnamefont {Nohara}}, \bibinfo {author} {\bibfnamefont
  {T.}~\bibnamefont {Fujita}}, \bibinfo {author} {\bibfnamefont
  {A.}~\bibnamefont {Mackenzie}}, \bibinfo {author} {\bibfnamefont
  {N.}~\bibnamefont {Hussey}}, \bibinfo {author} {\bibfnamefont
  {J.}~\bibnamefont {Bednorz}}, \ and\ \bibinfo {author} {\bibfnamefont
  {F.}~\bibnamefont {Lichtenberg}},\ }\href {\doibase 10.1143/JPSJ.66.1405}
  {\bibfield  {journal} {\bibinfo  {journal} {J. Phys. Soc. Jpn.}\ }\textbf
  {\bibinfo {volume} {66}},\ \bibinfo {pages} {1405} (\bibinfo {year}
  {1997})}\BibitemShut {NoStop}%
\bibitem [{\citenamefont {Urano}\ \emph {et~al.}(2000)\citenamefont {Urano},
  \citenamefont {Nohara}, \citenamefont {Kondo}, \citenamefont {Sakai},
  \citenamefont {Takagi}, \citenamefont {Shiraki},\ and\ \citenamefont
  {Okubo}}]{Urano2000}%
  \BibitemOpen
  \bibfield  {author} {\bibinfo {author} {\bibfnamefont {C.}~\bibnamefont
  {Urano}}, \bibinfo {author} {\bibfnamefont {M.}~\bibnamefont {Nohara}},
  \bibinfo {author} {\bibfnamefont {S.}~\bibnamefont {Kondo}}, \bibinfo
  {author} {\bibfnamefont {F.}~\bibnamefont {Sakai}}, \bibinfo {author}
  {\bibfnamefont {H.}~\bibnamefont {Takagi}}, \bibinfo {author} {\bibfnamefont
  {T.}~\bibnamefont {Shiraki}}, \ and\ \bibinfo {author} {\bibfnamefont
  {T.}~\bibnamefont {Okubo}},\ }\href {\doibase 10.1103/PhysRevLett.85.1052}
  {\bibfield  {journal} {\bibinfo  {journal} {Phys. Rev. Lett.}\ }\textbf
  {\bibinfo {volume} {85}},\ \bibinfo {pages} {1052} (\bibinfo {year}
  {2000})}\BibitemShut {NoStop}%
\bibitem [{\citenamefont {Miyake}\ \emph {et~al.}(1989)\citenamefont {Miyake},
  \citenamefont {Matsuura},\ and\ \citenamefont {Varma}}]{Miyake1989}%
  \BibitemOpen
  \bibfield  {author} {\bibinfo {author} {\bibfnamefont {K.}~\bibnamefont
  {Miyake}}, \bibinfo {author} {\bibfnamefont {T.}~\bibnamefont {Matsuura}}, \
  and\ \bibinfo {author} {\bibfnamefont {C.}~\bibnamefont {Varma}},\ }\href
  {\doibase https://doi.org/10.1016/0038-1098(89)90729-1} {\bibfield  {journal}
  {\bibinfo  {journal} {Solid State Commun.}\ }\textbf {\bibinfo {volume}
  {71}},\ \bibinfo {pages} {1149 } (\bibinfo {year} {1989})}\BibitemShut
  {NoStop}%
\bibitem [{\citenamefont {Zlati\'{c}}\ \emph {et~al.}(2003)\citenamefont
  {Zlati\'{c}}, \citenamefont {Horvati\'{c}}, \citenamefont {Milat},
  \citenamefont {Coqblin}, \citenamefont {Czycholl},\ and\ \citenamefont
  {Grenzebach}}]{Zlatic2003}%
  \BibitemOpen
  \bibfield  {author} {\bibinfo {author} {\bibfnamefont {V.}~\bibnamefont
  {Zlati\'{c}}}, \bibinfo {author} {\bibfnamefont {B.}~\bibnamefont
  {Horvati\'{c}}}, \bibinfo {author} {\bibfnamefont {I.}~\bibnamefont {Milat}},
  \bibinfo {author} {\bibfnamefont {B.}~\bibnamefont {Coqblin}}, \bibinfo
  {author} {\bibfnamefont {G.}~\bibnamefont {Czycholl}}, \ and\ \bibinfo
  {author} {\bibfnamefont {C.}~\bibnamefont {Grenzebach}},\ }\href {\doibase
  10.1103/PhysRevB.68.104432} {\bibfield  {journal} {\bibinfo  {journal} {Phys.
  Rev. B}\ }\textbf {\bibinfo {volume} {68}},\ \bibinfo {pages} {104432}
  (\bibinfo {year} {2003})}\BibitemShut {NoStop}%
\bibitem [{\citenamefont {Jaccard}\ \emph {et~al.}(1992)\citenamefont
  {Jaccard}, \citenamefont {Behnia},\ and\ \citenamefont
  {Sierro}}]{Jaccard1992}%
  \BibitemOpen
  \bibfield  {author} {\bibinfo {author} {\bibfnamefont {D.}~\bibnamefont
  {Jaccard}}, \bibinfo {author} {\bibfnamefont {K.}~\bibnamefont {Behnia}}, \
  and\ \bibinfo {author} {\bibfnamefont {J.}~\bibnamefont {Sierro}},\ }\href
  {http://www.sciencedirect.com/science/article/pii/037596019290860O}
  {\bibfield  {journal} {\bibinfo  {journal} {Phys. Lett. A}\ }\textbf
  {\bibinfo {volume} {163}},\ \bibinfo {pages} {475} (\bibinfo {year}
  {1992})}\BibitemShut {NoStop}%
\bibitem [{\citenamefont {Hodovanets}\ \emph {et~al.}(2015)\citenamefont
  {Hodovanets}, \citenamefont {Bud'ko}, \citenamefont {Straszheim},
  \citenamefont {Taufour}, \citenamefont {Mun}, \citenamefont {Kim},
  \citenamefont {Flint},\ and\ \citenamefont {Canfield}}]{Hodovanets2015}%
  \BibitemOpen
  \bibfield  {author} {\bibinfo {author} {\bibfnamefont {H.}~\bibnamefont
  {Hodovanets}}, \bibinfo {author} {\bibfnamefont {S.~L.}\ \bibnamefont
  {Bud'ko}}, \bibinfo {author} {\bibfnamefont {W.~E.}\ \bibnamefont
  {Straszheim}}, \bibinfo {author} {\bibfnamefont {V.}~\bibnamefont {Taufour}},
  \bibinfo {author} {\bibfnamefont {E.~D.}\ \bibnamefont {Mun}}, \bibinfo
  {author} {\bibfnamefont {H.}~\bibnamefont {Kim}}, \bibinfo {author}
  {\bibfnamefont {R.}~\bibnamefont {Flint}}, \ and\ \bibinfo {author}
  {\bibfnamefont {P.~C.}\ \bibnamefont {Canfield}},\ }\href {\doibase
  10.1103/PhysRevLett.114.236601} {\bibfield  {journal} {\bibinfo  {journal}
  {Phys. Rev. Lett.}\ }\textbf {\bibinfo {volume} {114}},\ \bibinfo {pages}
  {236601} (\bibinfo {year} {2015})}\BibitemShut {NoStop}%
\bibitem [{\citenamefont {Ren}\ \emph {et~al.}(2016)\citenamefont {Ren},
  \citenamefont {Scheerer}, \citenamefont {Lapertot},\ and\ \citenamefont
  {Jaccard}}]{Ren2016}%
  \BibitemOpen
  \bibfield  {author} {\bibinfo {author} {\bibfnamefont {Z.}~\bibnamefont
  {Ren}}, \bibinfo {author} {\bibfnamefont {G.~W.}\ \bibnamefont {Scheerer}},
  \bibinfo {author} {\bibfnamefont {G.}~\bibnamefont {Lapertot}}, \ and\
  \bibinfo {author} {\bibfnamefont {D.}~\bibnamefont {Jaccard}},\ }\href
  {\doibase 10.1103/PhysRevB.94.024522} {\bibfield  {journal} {\bibinfo
  {journal} {Phys. Rev. B}\ }\textbf {\bibinfo {volume} {94}},\ \bibinfo
  {pages} {024522} (\bibinfo {year} {2016})}\BibitemShut {NoStop}%
\bibitem [{\citenamefont {Fan}\ \emph {et~al.}(2004)\citenamefont {Fan},
  \citenamefont {Lee},\ and\ \citenamefont {Chen}}]{Fan2004}%
  \BibitemOpen
  \bibfield  {author} {\bibinfo {author} {\bibfnamefont {Y.~T.}\ \bibnamefont
  {Fan}}, \bibinfo {author} {\bibfnamefont {W.~H.}\ \bibnamefont {Lee}}, \ and\
  \bibinfo {author} {\bibfnamefont {Y.~Y.}\ \bibnamefont {Chen}},\ }\href
  {\doibase 10.1103/PhysRevB.69.132401} {\bibfield  {journal} {\bibinfo
  {journal} {Phys. Rev. B}\ }\textbf {\bibinfo {volume} {69}},\ \bibinfo
  {pages} {132401} (\bibinfo {year} {2004})}\BibitemShut {NoStop}%
\bibitem [{\citenamefont {Si}\ and\ \citenamefont {Steglich}(2010)}]{Si2010}%
  \BibitemOpen
  \bibfield  {author} {\bibinfo {author} {\bibfnamefont {Q.}~\bibnamefont
  {Si}}\ and\ \bibinfo {author} {\bibfnamefont {F.}~\bibnamefont {Steglich}},\
  }\href {\doibase 10.1126/science.1191195} {\bibfield  {journal} {\bibinfo
  {journal} {Science}\ }\textbf {\bibinfo {volume} {329}},\ \bibinfo {pages}
  {1161} (\bibinfo {year} {2010})}\BibitemShut {NoStop}%
\bibitem [{\citenamefont {Huo}\ \emph {et~al.}(2002)\citenamefont {Huo},
  \citenamefont {Sakurai}, \citenamefont {Kuwai}, \citenamefont {Mizushima},\
  and\ \citenamefont {Isikawa}}]{Huo2002}%
  \BibitemOpen
  \bibfield  {author} {\bibinfo {author} {\bibfnamefont {D.}~\bibnamefont
  {Huo}}, \bibinfo {author} {\bibfnamefont {J.}~\bibnamefont {Sakurai}},
  \bibinfo {author} {\bibfnamefont {T.}~\bibnamefont {Kuwai}}, \bibinfo
  {author} {\bibfnamefont {T.}~\bibnamefont {Mizushima}}, \ and\ \bibinfo
  {author} {\bibfnamefont {Y.}~\bibnamefont {Isikawa}},\ }\href {\doibase
  10.1103/PhysRevB.65.144450} {\bibfield  {journal} {\bibinfo  {journal} {Phys.
  Rev. B}\ }\textbf {\bibinfo {volume} {65}},\ \bibinfo {pages} {144450}
  (\bibinfo {year} {2002})}\BibitemShut {NoStop}%
\bibitem [{\citenamefont {Bediako}\ \emph {et~al.}(2018)\citenamefont
  {Bediako}, \citenamefont {Rezaee}, \citenamefont {Yoo}, \citenamefont
  {Larson}, \citenamefont {Zhao}, \citenamefont {Taniguchi}, \citenamefont
  {Watanabe}, \citenamefont {Brower-Thomas}, \citenamefont {Kaxiras},\ and\
  \citenamefont {Kim}}]{Bediako2018}%
  \BibitemOpen
  \bibfield  {author} {\bibinfo {author} {\bibfnamefont {D.~K.}\ \bibnamefont
  {Bediako}}, \bibinfo {author} {\bibfnamefont {M.}~\bibnamefont {Rezaee}},
  \bibinfo {author} {\bibfnamefont {H.}~\bibnamefont {Yoo}}, \bibinfo {author}
  {\bibfnamefont {D.~T.}\ \bibnamefont {Larson}}, \bibinfo {author}
  {\bibfnamefont {S.~Y.~F.}\ \bibnamefont {Zhao}}, \bibinfo {author}
  {\bibfnamefont {T.}~\bibnamefont {Taniguchi}}, \bibinfo {author}
  {\bibfnamefont {K.}~\bibnamefont {Watanabe}}, \bibinfo {author}
  {\bibfnamefont {T.~L.}\ \bibnamefont {Brower-Thomas}}, \bibinfo {author}
  {\bibfnamefont {E.}~\bibnamefont {Kaxiras}}, \ and\ \bibinfo {author}
  {\bibfnamefont {P.}~\bibnamefont {Kim}},\ }\href
  {https://doi.org/10.1038/s41586-018-0205-0} {\bibfield  {journal} {\bibinfo
  {journal} {Nature}\ }\textbf {\bibinfo {volume} {558}},\ \bibinfo {pages}
  {425} (\bibinfo {year} {2018})}\BibitemShut {NoStop}%
\end{thebibliography}

\begin{thebibliography}{7}%
\makeatletter
\providecommand \@ifxundefined [1]{%
 \@ifx{#1\undefined}
}%
\providecommand \@ifnum [1]{%
 \ifnum #1\expandafter \@firstoftwo
 \else \expandafter \@secondoftwo
 \fi
}%
\providecommand \@ifx [1]{%
 \ifx #1\expandafter \@firstoftwo
 \else \expandafter \@secondoftwo
 \fi
}%
\providecommand \natexlab [1]{#1}%
\providecommand \enquote  [1]{``#1''}%
\providecommand \bibnamefont  [1]{#1}%
\providecommand \bibfnamefont [1]{#1}%
\providecommand \citenamefont [1]{#1}%
\providecommand \href@noop [0]{\@secondoftwo}%
\providecommand \href [0]{\begingroup \@sanitize@url \@href}%
\providecommand \@href[1]{\@@startlink{#1}\@@href}%
\providecommand \@@href[1]{\endgroup#1\@@endlink}%
\providecommand \@sanitize@url [0]{\catcode `\\12\catcode `\$12\catcode
  `\&12\catcode `\#12\catcode `\^12\catcode `\_12\catcode `\%12\relax}%
\providecommand \@@startlink[1]{}%
\providecommand \@@endlink[0]{}%
\providecommand \url  [0]{\begingroup\@sanitize@url \@url }%
\providecommand \@url [1]{\endgroup\@href {#1}{\urlprefix }}%
\providecommand \urlprefix  [0]{URL }%
\providecommand \Eprint [0]{\href }%
\providecommand \doibase [0]{http://dx.doi.org/}%
\providecommand \selectlanguage [0]{\@gobble}%
\providecommand \bibinfo  [0]{\@secondoftwo}%
\providecommand \bibfield  [0]{\@secondoftwo}%
\providecommand \translation [1]{[#1]}%
\providecommand \BibitemOpen [0]{}%
\providecommand \bibitemStop [0]{}%
\providecommand \bibitemNoStop [0]{.\EOS\space}%
\providecommand \EOS [0]{\spacefactor3000\relax}%
\providecommand \BibitemShut  [1]{\csname bibitem#1\endcsname}%
\let\auto@bib@innerbib\@empty
%</preamble>
\bibitem [{\citenamefont {Niu}\ \emph {et~al.}(2017)\citenamefont {Niu},
  \citenamefont {Yan}, \citenamefont {Ji}, \citenamefont {Liu}, \citenamefont
  {Li}, \citenamefont {Gao}, \citenamefont {Zhang}, \citenamefont {Yu},\ and\
  \citenamefont {Wu}}]{Niu2017}%
  \BibitemOpen
  \bibfield  {author} {\bibinfo {author} {\bibfnamefont {J.}~\bibnamefont
  {Niu}}, \bibinfo {author} {\bibfnamefont {B.}~\bibnamefont {Yan}}, \bibinfo
  {author} {\bibfnamefont {Q.}~\bibnamefont {Ji}}, \bibinfo {author}
  {\bibfnamefont {Z.}~\bibnamefont {Liu}}, \bibinfo {author} {\bibfnamefont
  {M.}~\bibnamefont {Li}}, \bibinfo {author} {\bibfnamefont {P.}~\bibnamefont
  {Gao}}, \bibinfo {author} {\bibfnamefont {Y.}~\bibnamefont {Zhang}}, \bibinfo
  {author} {\bibfnamefont {D.}~\bibnamefont {Yu}}, \ and\ \bibinfo {author}
  {\bibfnamefont {X.}~\bibnamefont {Wu}},\ }\href {\doibase
  10.1103/PhysRevB.96.075402} {\bibfield  {journal} {\bibinfo  {journal} {Phys.
  Rev. B}\ }\textbf {\bibinfo {volume} {96}},\ \bibinfo {pages} {075402}
  (\bibinfo {year} {2017})}\BibitemShut {NoStop}%
\bibitem [{\citenamefont {Schlottmann}(1989)}]{Schlottmann1989a}%
  \BibitemOpen
  \bibfield  {author} {\bibinfo {author} {\bibfnamefont {P.}~\bibnamefont
  {Schlottmann}},\ }\href
  {http://www.sciencedirect.com/science/article/pii/0370157389901166}
  {\bibfield  {journal} {\bibinfo  {journal} {Phys. Rep.}\ }\textbf {\bibinfo
  {volume} {181}},\ \bibinfo {pages} {1} (\bibinfo {year} {1989})}\BibitemShut
  {NoStop}%
\bibitem [{\citenamefont {Hossain}\ \emph {et~al.}(2000)\citenamefont
  {Hossain}, \citenamefont {Hamashima}, \citenamefont {Umeo}, \citenamefont
  {Takabatake}, \citenamefont {Geibel},\ and\ \citenamefont
  {Steglich}}]{Hossain2000a}%
  \BibitemOpen
  \bibfield  {author} {\bibinfo {author} {\bibfnamefont {Z.}~\bibnamefont
  {Hossain}}, \bibinfo {author} {\bibfnamefont {S.}~\bibnamefont {Hamashima}},
  \bibinfo {author} {\bibfnamefont {K.}~\bibnamefont {Umeo}}, \bibinfo {author}
  {\bibfnamefont {T.}~\bibnamefont {Takabatake}}, \bibinfo {author}
  {\bibfnamefont {C.}~\bibnamefont {Geibel}}, \ and\ \bibinfo {author}
  {\bibfnamefont {F.}~\bibnamefont {Steglich}},\ }\href {\doibase
  10.1103/PhysRevB.62.8950} {\bibfield  {journal} {\bibinfo  {journal} {Phys.
  Rev. B}\ }\textbf {\bibinfo {volume} {62}},\ \bibinfo {pages} {8950}
  (\bibinfo {year} {2000})}\BibitemShut {NoStop}%
\bibitem [{\citenamefont {Pikul}\ \emph {et~al.}(2003)\citenamefont {Pikul},
  \citenamefont {Kaczorowski}, \citenamefont {Plackowski}, \citenamefont
  {Czopnik}, \citenamefont {Michor}, \citenamefont {Bauer}, \citenamefont
  {Hilscher}, \citenamefont {Rogl},\ and\ \citenamefont {Grin}}]{Pikul2003a}%
  \BibitemOpen
  \bibfield  {author} {\bibinfo {author} {\bibfnamefont {A.~P.}\ \bibnamefont
  {Pikul}}, \bibinfo {author} {\bibfnamefont {D.}~\bibnamefont {Kaczorowski}},
  \bibinfo {author} {\bibfnamefont {T.}~\bibnamefont {Plackowski}}, \bibinfo
  {author} {\bibfnamefont {A.}~\bibnamefont {Czopnik}}, \bibinfo {author}
  {\bibfnamefont {H.}~\bibnamefont {Michor}}, \bibinfo {author} {\bibfnamefont
  {E.}~\bibnamefont {Bauer}}, \bibinfo {author} {\bibfnamefont
  {G.}~\bibnamefont {Hilscher}}, \bibinfo {author} {\bibfnamefont
  {P.}~\bibnamefont {Rogl}}, \ and\ \bibinfo {author} {\bibfnamefont
  {Y.}~\bibnamefont {Grin}},\ }\href {\doibase 10.1103/PhysRevB.67.224417}
  {\bibfield  {journal} {\bibinfo  {journal} {Phys. Rev. B}\ }\textbf {\bibinfo
  {volume} {67}},\ \bibinfo {pages} {224417} (\bibinfo {year}
  {2003})}\BibitemShut {NoStop}%
\bibitem [{\citenamefont {Zhou}\ \emph {et~al.}(2018)\citenamefont {Zhou},
  \citenamefont {Xu}, \citenamefont {Li}, \citenamefont {Sankar}, \citenamefont
  {Zhang}, \citenamefont {Qian}, \citenamefont {Cao}, \citenamefont {Dai},
  \citenamefont {Lu},\ and\ \citenamefont {Xu}}]{Zhou2018b}%
  \BibitemOpen
  \bibfield  {author} {\bibinfo {author} {\bibfnamefont {W.}~\bibnamefont
  {Zhou}}, \bibinfo {author} {\bibfnamefont {C.~Q.}\ \bibnamefont {Xu}},
  \bibinfo {author} {\bibfnamefont {B.}~\bibnamefont {Li}}, \bibinfo {author}
  {\bibfnamefont {R.}~\bibnamefont {Sankar}}, \bibinfo {author} {\bibfnamefont
  {F.~M.}\ \bibnamefont {Zhang}}, \bibinfo {author} {\bibfnamefont
  {B.}~\bibnamefont {Qian}}, \bibinfo {author} {\bibfnamefont {C.}~\bibnamefont
  {Cao}}, \bibinfo {author} {\bibfnamefont {J.~H.}\ \bibnamefont {Dai}},
  \bibinfo {author} {\bibfnamefont {J.}~\bibnamefont {Lu}}, \ and\ \bibinfo
  {author} {\bibfnamefont {X.}~\bibnamefont {Xu}},\ }\href@noop {} {\bibfield
  {journal} {\bibinfo  {journal} {ArXiv e-prints}\ ,\ \bibinfo {pages}
  {1802.05060}} (\bibinfo {year} {2018})}\BibitemShut {NoStop}%
\bibitem [{\citenamefont {Perdew}\ \emph {et~al.}(1996)\citenamefont {Perdew},
  \citenamefont {Burke},\ and\ \citenamefont {Ernzerhof}}]{Perdew1996a}%
  \BibitemOpen
  \bibfield  {author} {\bibinfo {author} {\bibfnamefont {J.~P.}\ \bibnamefont
  {Perdew}}, \bibinfo {author} {\bibfnamefont {K.}~\bibnamefont {Burke}}, \
  and\ \bibinfo {author} {\bibfnamefont {M.}~\bibnamefont {Ernzerhof}},\ }\href
  {\doibase 10.1103/PhysRevLett.77.3865} {\bibfield  {journal} {\bibinfo
  {journal} {Phys. Rev. Lett.}\ }\textbf {\bibinfo {volume} {77}},\ \bibinfo
  {pages} {3865} (\bibinfo {year} {1996})}\BibitemShut {NoStop}%
\bibitem [{\citenamefont {Blum}\ \emph {et~al.}(2009)\citenamefont {Blum},
  \citenamefont {Gehrke}, \citenamefont {Hanke}, \citenamefont {Havu},
  \citenamefont {Havu}, \citenamefont {Ren}, \citenamefont {Reuter},\ and\
  \citenamefont {Scheffler}}]{Blum2009a}%
  \BibitemOpen
  \bibfield  {author} {\bibinfo {author} {\bibfnamefont {V.}~\bibnamefont
  {Blum}}, \bibinfo {author} {\bibfnamefont {R.}~\bibnamefont {Gehrke}},
  \bibinfo {author} {\bibfnamefont {F.}~\bibnamefont {Hanke}}, \bibinfo
  {author} {\bibfnamefont {P.}~\bibnamefont {Havu}}, \bibinfo {author}
  {\bibfnamefont {V.}~\bibnamefont {Havu}}, \bibinfo {author} {\bibfnamefont
  {X.~G.}\ \bibnamefont {Ren}}, \bibinfo {author} {\bibfnamefont
  {K.}~\bibnamefont {Reuter}}, \ and\ \bibinfo {author} {\bibfnamefont
  {M.}~\bibnamefont {Scheffler}},\ }\href {\doibase 10.1016/j.cpc.2009.06.022}
  {\bibfield  {journal} {\bibinfo  {journal} {Comput. Phys. Commun.}\ }\textbf
  {\bibinfo {volume} {180}},\ \bibinfo {pages} {2175} (\bibinfo {year}
  {2009})}\BibitemShut {NoStop}%
\end{thebibliography}
%

\clearpage

\end{document}